 \documentstyle[eqsecnum,prb,aps]{revtex}

\begin{document}


\draft
\preprint{}
\title{Vortex lattice structure in a $d_{x^2-y^2}$-wave superconductor} 
\author{Masanori Ichioka, Naoki Enomoto, and Kazushige Machida}
\address{Department of Physics, Okayama University, Okayama 700, Japan}
\date{\today}
\maketitle
\begin{abstract}
The vortex lattice structure in a $d_{x^2-y^2}$-wave superconductor 
is investigated near the upper critical magnetic field in the framework 
of the Ginzburg Landau theory extended by including the 
correction terms such as the higher order derivatives 
derived from the Gor'kov equation. 
On lowering temperature, the unit cell shape of the vortex lattice 
gradually varies from a regular triangular lattice to a square lattice through 
the shape of an isosceles triangle. 
As for the orientation of the vortex lattice, the base of an isosceles 
triangle is along the $a$ axis or the $b$ axis of the crystal. 
The fourfold symmetric structure around a vortex core is also studied in 
the vortex lattice case. 
It is noted that these characteristic features appear even in the case 
the induced $s$-wave order parameter is absent around the vortex of the 
$d_{x^2-y^2}$-wave superconductivity. 
We also investigate the effect of the induced $s$-wave order parameter. 
It enhances (suppresses) these characteristic features of the 
$d_{x^2-y^2}$-wave superconductor when the $s$-wave component of the 
interaction is attractive (repulsive). 
\end{abstract}

\widetext

\section{Introduction}
\label{sec:1}

Much attention has been focused on a vortex structure in 
a high-$T_{\rm c}$  superconductor.
It is of great interest how the vortex structure of the 
high-$T_{\rm c}$ superconductor is different from that of 
a conventional superconductor. 
Various investigations of the flux line lattice (or vortex lattice) 
for conventional superconductors are compiled in Ref. \onlinecite{Weber}, 
including the gap and Fermi surface anisotropy effects. 
By a number of experimental and theoretical investigations, 
it is concluded that the symmetry of this superconductor is 
most likely to be  $d_{x^2-y^2}$-wave. 
Therefore, one of the points is how the vortex structure of 
a $d_{x^2-y^2}$-wave superconductor is different from that of 
an isotropic $s$-wave superconductor. 
The amplitude of the $d_{x^2-y^2}$-wave pairing, i.e., 
$\hat k_x^2 - \hat k_y^2$ in momentum space, has fourfold symmetry 
for the rotation about the $c$-axis. 
We expect that, reflecting this symmetry, the core structure of an 
isolated single vortex may break the cylindrical symmetry and show fourfold 
symmetry in a $d_{x^2-y^2}$-wave superconductor, 
and the fourfold symmetry of the core structure may affect the 
vortex lattice structure. 

Keimer {\it et al.}~\cite{Keimer} observed an oblique
lattice by a small-angle neutron scattering (SANS) study of the vortex lattice
in ${\rm YBa_2Cu_3O_7}$ in a magnetic field region of 0.5T$\le H \le$5T 
applied parallel to the $c$ axis.
They reported that the vortex lattice has an angle of 
$73^\circ$ between the two primitive
vectors and is oriented such that the nearest-neighbor direction of vortices
makes an angle of $45^\circ$ with the $a$ axis.
The oblique lattice with $77^\circ$ was also found by scanning 
tunneling microscopy (STM) at 6T by Maggio-Aprile 
{\it et al.}~\cite{Maggio,Renner} 
They observed the elliptic-shaped STM image of the vortex core, 
and concluded that this oblique lattice cannot be explained by considering 
only the effect of the intrinsic in-plane anisotropy, that is,
the difference of the coherence lengths between the $a$ axis and
the $b$ axis directions.~\cite{Renner,Walker}  
On the other hand, at low fields of 35 and 65G, 
Dolan {\it et al.}\cite{Dolan} reported the regular triangular lattice 
by the Bitter-pattern observation of the vortices, where the slight 
distortion of the vortex lattice is explained by the anisotropy 
of the coherence lengths between the $a$ axis and the $b$ axis directions. 
It is suggested that this deformation from the regular triangular lattice 
in the high field region of a $d_{x^2-y^2}$-wave superconductor is 
due to the effect of the fourfold symmetric vortex core structure. 

The fourfold symmetric vortex core structure in a $d_{x^2-y^2}$-wave 
superconductor was so far derived theoretically when the $s$-wave component 
is induced around a vortex of $d_{x^2-y^2}$-wave 
superconductivity.~\cite{Berlinsky,Franz,Affleck,Ren,Xu,RenB,Soininen,IchiokaS} 
This mixing scenario was mainly put forth  by Berlinsky
{\it et al.}\cite{Berlinsky,Franz} and Ren {\it et al.}~\cite{Ren,Xu,RenB} 
in the two-component Ginzburg Landau (GL) theory for $s$- and $d$-wave 
superconductivity.
According to the consideration based on the two-component GL theory,
it is possible that the $s$-wave component is coupled with the $d$-wave
component through the gradient terms.
Therefore, the $s$-wave component may be  induced when the $d$-wave
order parameter spatially varies, such as near the vortex or interface
under certain restricted conditions.\cite{Xu,Matsumoto}
The induced $s$-wave component around the vortex is fourfold symmetric.
The resulting vortex structure in the $d$-wave superconductor, therefore,
exhibits fourfold symmetry.

However, it is noted that the amplitude of the induced $s$-wave 
order parameter strongly depends on the detail of the pairing interaction. 
In the weak coupling theory of the continuum model, for example, 
we define that the $s$-wave component of the order parameter is   
proportional to $V_s$, the $s$-wave component of the pairing 
interaction.~\cite{IchiokaS} 
Therefore, when $|V_s|$ is negligibly small compared with the dominant 
$d_{x^2-y^2}$-wave component of the pairing interaction, 
the induced $s$-wave order parameter can be neglected. 
On the other hand, it was reported that 
in the Bogoliubov-de Gennes theory of the extended Hubbard model 
the amplitude of the induced $s$-wave order parameter is suppressed  
when the on-site repulsion of the interaction increases.~\cite{Soininen}
In the numerical calculation by Wang and MacDonald,~\cite{Wang} 	 
the induced $s$-wave order parameter is small, on the order of a 
few percents of the dominant $d_{x^2-y^2}$-wave component.

For a moment, we consider the limit where the induced $s$-wave 
order parameter is negligibly small compared with 
the dominant $d_{x^2-y^2}$-wave component.  
(We denote this limit as ``pure $d_{x^2-y^2}$-wave''.) 
In this case, the two-component GL equations are reduced to a 
one-component GL equation for only a $d_{x^2-y^2}$-wave component, 
which is the same form as that of the isotropic $s$-wave case 
within the conventional GL theory. 
Thus, the conventional GL theory gives the result that 
the vortex structure in a pure $d_{x^2-y^2}$-wave superconductor 
is exactly the same as that of an isotropic $s$-wave one, that is, 
the core structure of a single vortex is circular symmetric and 
the vortex lattice is the triangular Abrikosov lattice 
structure.~\cite{Abrikosov}

However, Ichioka {\it et al.}~\cite{IchiokaF} showed that the core 
structure of a single vortex exhibits the fourfold symmetric 
structure even in a pure $d_{x^2-y^2}$-wave superconductor 
by a numerical calculation based on the quasi-classical Eilenberger 
theory, which can be applied even at low temperatures. 
In their calculation, the fourfold symmetric structure becomes clear 
on lowering temperature $T$. 
These facts indicate the following results. 
Strictly speaking, the conventional GL equation is valid 
only near the transition temperature $T_c$. 
Far from $T_c$, we have to include several correction terms which is 
higher order in the small parameter $\ln(T_c/T)\simeq 1-T_c/T$. 
Enomoto {\it et al.}~\cite{Enomoto} showed that the correction terms 
consist of higher order terms of the order parameter and the higher 
derivatives (or non-local terms) which are neglected in the conventional 
GL theory. 
And they indeed succeeded in showing that the fourfold symmetric core 
structure of a single vortex in a pure $d_{x^2-y^2}$-wave superconductor 
is reproduced by including these correction terms. 
This indicates that, in the study of the fourfold symmetric vortex 
core structure in a $d_{x^2-y^2}$-wave superconductor, 
we have to consider the effect of the correction terms of the order 
$\ln(T_c/T)$ in addition to the effect of the induced $s$-wave 
order parameter. 
These correction terms may affect the structure of the vortex lattice. 
 
As for the effect of the induced $s$-wave order parameter on the 
vortex lattice structure,  
Berlinsky {\it et al.}~\cite{Berlinsky,Franz,Affleck} 
investigated it by using the two-component GL equations. 
As characteristic features of the $d_{x^2-y^2}$-wave superconductor's 
vortex lattice, they suggested that the vortex lattice is deformed 
from a triangular lattice due to the induced $s$-wave order parameter 
and it leads to the double peak structure in the magnetic field 
distribution function $P(h)$ as a function of magnetic field $h$.   
In a pure $d_{x^2-y^2}$-wave superconductor, however, 
these features vanish within their theory 
since the induced $s$-wave order parameter is absent.  
On the other hand, the correction terms which introduce the fourfold 
symmetric core structure might give some effects on the vortex lattice 
structure  even in the pure $d_{x^2-y^2}$-wave superconductor. 
In the study of the vortex lattice based on the GL framework, 
the effect of the correction terms in a $d_{x^2-y^2}$-wave 
superconductor has not been studied sufficiently. 

The purpose of this paper is to determine the orientation and the unit 
cell shape of the vortex lattice in the framework of the extended GL theory 
near the upper critical magnetic field $H_{c2}$. 
We also investigate the spatial variation of the vortex lattice structure, 
that is, the order parameter, the current and the induced magnetic field. 
In addition to the effect of the correction terms of the $d_{x^2-y^2}$-wave 
superconductivity in the order $\ln(T_c/T)$, 
our calculation is performed by including also the effect of the 
induced $s$-wave component so that the contribution of both effects 
(the effects of the correction terms and of the induced $s$-wave component) 
can be estimated on the equal footing. 
Therefore, we reconstruct the two-component GL theory for $s$- and 
$d_{x^2-y^2}$-wave order parameters with including the correction terms 
of the order $\ln(T_c/T)$. 
The pure $d_{x^2-y^2}$-wave case is easily obtained in the limit where 
the amplitude of the $s$-wave order parameter is negligibly small. 

We consider the case of the vortex lattice under a magnetic field 
applied parallel to the $c$ axis (or $z$ axis) in the clean limit. 
Our consideration is restricted in the case of tetragonal symmetry 
described by the point group $D_4$. 
Then, the $a$ axis and the $b$ axis of the crystal coordinate 
are equivalent each other. 
The orthorhombic symmetry case can be obtained by scaling the coordinate 
system, namely, the $a$ axis direction and the $b$ axis direction are made  
to be equivalent if the lengths are scaled by the coherence length 
of each direction. 
The Fermi surface is assumed to be two-dimensional, which is appropriate 
to high-$T_c$ superconductors, and isotropic in order to clarify effects 
of the $d_{x^2-y^2}$-wave nature of the pair potential on the vortex 
lattice structure. 
The additional anisotropy coming from, e.g., the Fermi surface can be 
incorporated into our extended GL framework. 

The two-component GL equations were derived, for example, by 
Ren {\it et al.}~\cite{Xu,RenB} from the Gor'kov equation. 
However, their derivation is within the conventional GL framework. 
Here, to study the effect of correction terms of the order $\ln(T_c/T)$, 
we reconstruct the GL theory by including higher order terms of 
the order parameter and the higher derivatives other than terms  
in the conventional GL theory. 
In the pure $d_{x^2-y^2}$-wave case, 
the present authors derived the GL equation with including these correction 
terms from the Gor'kov equation for the continuum model 
to study the fourfold symmetric core structure of a single 
vortex.~\cite{Enomoto}
In this paper, we derive the two-component GL equation for $s$- and 
$d_{x^2-y^2}$-wave components from the Gor'kov equation with including 
the correction terms of the order $\ln(T_c/T)$. 
For simplicity of calculation, our derivation is performed in the 
weak coupling theory of the continuum model. 
As for other models, for example, Feder and Kallin derived 
the two-component GL equations from the extended Hubbard model 
with including these higher order terms.~\cite{Feder}   
In the GL equation derived from other model, the coefficient of each 
term is modified, but the derived terms have one-to-one correspondence 
to those of ours. 
Therefore, while our results may be modified quantitatively, 
our approach can be applied to the GL equation derived from other models 
and our results remain to be valid qualitatively.

We calculate the $d_{x^2-y^2}$-wave and the induced $s$-wave components  
of the order parameter near $H_{c2}$ by solving the linearized version 
of the derived GL equations. 
We also calculate the current and the induced magnetic field 
in the vortex lattice near $H_{c2}$ by using the current density 
expression which includes the contribution of the correction terms 
and the induced $s$-wave order parameter. 
In our calculation of the vortex lattice structure, we use the 
magnetic Bloch state (or Eilenberger's basis function)~\cite{Eilenberger} 
as the basis function and expand various quantities 
in powers of $\ln(T_c/T)$. 
Then, we can determine the vortex lattice structure in terms of an 
analytically expressed form, which makes it easy to understand the detail 
of the properties of the vortex lattice, such as the dependence on the 
mixing of the $s$-wave component.

As for the orientation and the unit cell shape of the vortex lattice, 
the equilibrium  state is estimated by the minimum of the free energy. 
This was done by Abrikosov near $H_{c2}$ within the conventional GL theory 
for an isotropic $s$-wave superconductor.~\cite{Abrikosov}   
Here, we have to extend his theory so that we can  include the effects of 
the correction terms and of the induced $s$-wave order parameter. 
The extension about the effect of the induced $s$-wave order 
parameter was done by Berlinsky {\it et al.}~\cite{Berlinsky,Franz} 
The extension about the effect of the correction terms was performed 
by Takanaka in the case of isotropic $s$-wave pairing but anisotropic 
density of states at the Fermi surface.~\cite{TakanakaPTP} 
In this paper, we follow the method provided by Takanaka and 
extend his calculation to the case where the superconductivity 
is $d_{x^2-y^2}$-wave symmetry and the induced $s$-wave order 
parameter also exists. 

On the other hand, Won and Maki~\cite{WonMakiPRB,WonMaki} studied 
the vortex lattice structure near $H_{c2}$ numerically from 
the Gor'kov equation on the same stand point of ours. 
They suggested that the most stable configuration of the vortex 
lattice is a square lattice tilted by $\pi /4$ from the $a$ axis at 
temperature $T \le 0.88 T_c$ even in the pure $d_{x^2-y^2}$-wave case. 
In our interpretation, this deformation of the vortex lattice 
from a triangular lattice is due to the above-mentioned correction 
terms of the order $\ln(T_c/T)$, which are derived from the Gor'kov 
equation but neglected in the conventional GL theory. 
Our calculation is based on the GL expansion of the Gor'kov equation. 
Therefore, their numerical results are reproduced near $T_c$ 
by our calculation as an analytically expressed form.  
We note that Won and Maki made some errors in their estimate  
of the free energy minimum. 
They estimated the free energy minimum by the Abrikosov parameter 
$\beta_A$ which is originally used within the conventional GL  theory.  
However, this expression of $\beta_A$ itself should be modified by 
including the correction of the order $\ln(T_c/T)$. 
In their calculation, the free energy is estimated only in 
the case of a regular triangular lattice and a square lattice. 
In addition to them, the oblique lattice should be considered together. 
In our calculation, we derive the correct expression of 
the Abrikosov parameter. 
And, by using it, we estimate the free energy minimum by exhausting 
all the possible orientation and unit cell shape of the vortex lattice 
structure, including a shape of a scalene triangle. 

The rest of this paper is organized as follows. 
In Sec. \ref{sec:2}, we construct the two-component GL equations 
with including the correction terms of the order $\ln(T_c/T)$. 
By using these equations, the $d_{x^2-y^2}$-wave and the induced 
$s$-wave component of the order parameter are calculated near $H_{c2}$. 
In Sec. \ref{sec:3}, we calculate the current and the induced magnetic 
field in the vortex lattice state near $H_{c2}$. 
The orientation and the unit cell shape of the equilibrium vortex 
lattice structure is determined by the free energy minimum 
in Sec. \ref{sec:4}. 
In Sec. \ref{sec:5}, we present the spatial variation of the vortex 
lattice structure in the pure $d_{x^2-y^2}$-wave case.  
The summary and discussions are given in Sec. \ref{sec:6}.

\section{Order Parameter near $H_{c2}$}
\label{sec:2} 

We adopt the coordinate system shown in Fig. \ref{fig:1}, where one of 
the unit vectors of the vortex lattice is along the $x$ axis. 
Thus, the unit vectors of the vortex lattice are given by 
${\bf r}_1=(a_x,0)$ and ${\bf r}_2=(\zeta a_x, a_y)$. 
The $x$ axis forms an angle of $\theta_0$ from the $a$ axis of 
the crystal coordinate  as shown in Fig. \ref{fig:1}.  
Therefore the shape of the vortex lattice is characterized by 
$a_y/a_x$ and $\zeta$, and the orientation of the vortex lattice 
is denoted by $\theta_0$. 

We consider the two-component GL theory for $s$- and 
$d_{x^2-y^2}$-wave superconductivity. 
The order parameter (or pair potential) is given by 
\begin{equation}
\Delta({\bf r},\theta)
=s({\bf r})\phi_s(\theta)+d({\bf r})\phi_d(\theta) . 
\label{eq:2.1}
\end{equation}
Here, ${\bf r}=(x,y)$ is the center of mass coordinate of the 
Cooper pair.
The direction of relative momentum of the Cooper pair is denoted 
by an angle $\theta$ measured from the $x$ axis in the $ab$ plane. 
The symmetry function for $s$-wave ($d_{x^2-y^2}$-wave) component 
$\phi_s(\theta)$ ($\phi_d(\theta)$) is assumed to be real and 
to satisfy the normalization condition $\langle \phi_s^2 \rangle =1$
($\langle \phi_d^2 \rangle =1$). 
Here, we write $\int( \cdots )d\theta /2 \pi =\langle\cdots\rangle$. 
In our presentation of figures, we use forms 
\begin{equation}
\phi_s(\theta)=1, \quad 
\phi_d(\theta)=\sqrt{2}\cos 2(\theta-\theta_0) 
\label{eq:2.2}
\end{equation}
as an example. 
On the other hand, the pairing interaction is assumed to 
be separable as follows, 
\begin{equation}
V(\theta',\theta)=V_s \phi_s(\theta')\phi_s(\theta) 
+V_d \phi_d(\theta')\phi_d(\theta) , 
\label{eq:2.3}
\end{equation}
where attractive interaction is treated as positive. 
In Eq. (\ref{eq:2.3}), the $d_{x^2-y^2}$-wave component $V_d$ is dominant 
and the $s$-wave component $V_s$ is smaller than $V_d$.

The GL equation is derived as follows from the Gor'kov equation 
in the weak coupling theory of the continuum model 
(As for the detail of the derivation, see Appendix A of 
Ref. \onlinecite{Enomoto}. 
Now, $\eta({\bf r})\phi({\bf k})$ and 
$\bar{V}\phi^\ast({\bf k})\phi({\bf k}')$ are denoted by
$\Delta({\bf r},\theta)$ and $V(\theta,\theta')$, respectively.), 
\widetext 
\begin{eqnarray} 
N_F^{-1}\Delta({\bf r},\theta')
&=&
A_0 \langle V(\theta',\theta)\Delta({\bf r},\theta) \rangle
+A_2 v_F^2 \langle V(\theta',\theta)
(\hat{\bf v}\cdot {\bf q})^2 \Delta({\bf r},\theta) \rangle
+A_2 \langle V(\theta',\theta)
|\Delta({\bf r},\theta)|^2 \Delta({\bf r},\theta) \rangle
\nonumber \\ 
&+& 
A_4 v_F^4 \langle V(\theta',\theta)
(\hat{\bf v}\cdot {\bf q})^4 \Delta({\bf r},\theta) \rangle
+6A_4 \langle V(\theta',\theta)|\Delta({\bf r},\theta)|^4
\Delta({\bf r},\theta) \rangle
\nonumber \\ 
&+& 
2A_4 v_F^2 \langle V(\theta',\theta) [ 
4 \{ (\hat{\bf v}\cdot {\bf q})^2 \Delta({\bf r},\theta) \}
|\Delta({\bf r},\theta)|^2 
+\Delta({\bf r},\theta)^2 
\{ (\hat{\bf v}\cdot {\bf q})^2 \Delta({\bf r},\theta) \}^\ast 
\nonumber \\ &&
-2|(\hat{\bf v}\cdot {\bf q})\Delta({\bf r},\theta)|^2 
\Delta({\bf r},\theta) 
+3 \{ (\hat{\bf v}\cdot {\bf q})\Delta({\bf r},\theta) \}^2
\Delta^\ast({\bf r},\theta) ]\rangle
+ ({\rm higher \ order \ terms})  
\label{eq:2.4}
\end{eqnarray}
with $\hat{\bf v} \equiv {\bf v}(\theta)/v_F
=(\cos\theta,\sin\theta)$ and 
${\bf q}=  \nabla / i + (2\pi/\phi_0){\bf A}$, 
where $N_F$ is the density of states at the Fermi surface,  
$v_F$ the Fermi velocity and $\phi_0$ the flux quantum. 
The vector potential is given by ${\bf A}=(-H_0 y,0,0)$ in the 
Landau gauge. 
When we consider the GL equation, the contribution of the internal 
magnetic field can be neglected in ${\bf A}$ compared with that of 
the external field $H_0$, since we consider the case of 
an extreme type II superconductor. 
In Eq. (\ref{eq:2.4}), $A_0=\ln(2\omega_D \bar{\gamma}/\pi T)$ 
with the Debye frequency $\omega_D$ and the Euler constant 
$\bar{\gamma}$, $A_2=-\beta/3$ and $A_4=\alpha \beta^2 /18$, where 
\begin{equation}
\alpha=\frac{62\zeta(5)}{49\zeta(3)^2}=0.908..., \qquad
\beta= \frac{21\zeta(3)}{16\pi^2T^2} 
\label{eq:2.5}
\end{equation}
with Riemann's $\zeta$-functions $\zeta(3)$ and $\zeta(5)$.
The terms with $A_4$ in Eq. (\ref{eq:2.4}) are the correction terms 
which are neglected within the conventional GL theory. 

Substituting Eqs. (\ref{eq:2.1}) and (\ref{eq:2.3}) into 
Eq. (\ref{eq:2.4}) and using the relation 
$(N_FV_d)^{-1}
=\ln(2\omega_{\rm D}\bar{\gamma} / \pi T_c )$, we obtain 
two-component GL equations for $s$- and $d_{x^2-y^2}$-wave 
components, 
\begin{eqnarray} && 
\frac{1}{\eta_0}s+ \gamma c_s \frac{2 \xi^2}{\eta_0} 
\left\{ 
  \langle \phi_s^2 (\hat{\bf v}\cdot {\bf q})^2 \rangle s
+ \langle \phi_s \phi_d (\hat{\bf v}\cdot {\bf q})^2 \rangle d 
\right\}
+\gamma c_s \frac{2}{3 \eta_0^3}
\left\{ 
  \langle \phi_s^4 \rangle |s|^2 s 
 +\langle \phi_s^2 \phi_d^2 \rangle (2|d|^2 s + d^2 s^\ast) 
\right\}
+ O(\gamma^2)=0, 
\label{eq:2.6}
\end{eqnarray}
\begin{eqnarray}&&
-\frac{1}{\eta_0}d + \frac{2 \xi^2}{\eta_0}
\left\{ 
  \langle \phi_d^2 (\hat{\bf v}\cdot {\bf q})^2 \rangle d
+ \langle \phi_s \phi_d (\hat{\bf v}\cdot {\bf q})^2 \rangle s 
\right\}
+\frac{2}{3 \eta_0^3}
\left\{ 
  \langle \phi_d^4 \rangle |d|^2 d
 +\langle \phi_s^2 \phi_d^2 \rangle (2|s|^2 d + s^2 d^\ast ) 
\right\}
\nonumber \\ &&
-\gamma \Bigl[
\frac{2 \xi^4}{\eta_0} 
\langle \phi_d^2 (\hat{\bf v}\cdot {\bf q})^4 \rangle d 
+\frac{1}{3 \eta_0^5} \langle \phi_d^6 \rangle |d|^4 d 
-\frac{2 \xi^2}{3 \eta_0^3 } \langle \phi_d^4 [ 
4 \{(\hat{\bf v}\cdot {\bf q})^2 d \} |d|^2
+d^2 \{ (\hat{\bf v}\cdot {\bf q})^2 d \}^\ast
\nonumber \\ &&
-2d|(\hat{\bf v}\cdot {\bf q})d|^2 
+ 3 \{ (\hat{\bf v}\cdot {\bf q})d \} ^2d^\ast ] \rangle \Bigr]
+ O(\gamma^2)=0,
\label{eq:2.7}
\end{eqnarray}
with $\gamma = \alpha \ln(T_c/T)$, 
the GL coherence length 
$\xi=\{\beta v_F^2 /6 \ln(T_{\rm c}/T)\}^{1/2}$  
and the energy gap in the absence of a magnetic field 
$\eta_0=\{\ln(T_{\rm c}/T)/\beta \}^{1/2}$. 
Because $s({\bf r})$ is in the order $O(\gamma)$ as shown later 
in Eq. (\ref{eq:2.26}), the terms including $s({\bf r})$ are neglected 
among the correction terms of the order $O(\gamma)$ in Eq. (\ref{eq:2.7}).
The parameter $c_s$ in Eq. (\ref{eq:2.6}) is defined by
\begin{eqnarray} 
c_s^{-1} &\equiv& \alpha\{ (N_F V_s)^{-1} -(N_F V_d)^{-1}-\ln(T_c/T) \} 
=
\alpha\{ (N_F V_s)^{-1}-(N_F V_d)^{-1}\} +O(\gamma), 
\label{eq:2.8}
\end{eqnarray}
which depends on the interaction parameters $V_s$ and $V_d$.
From Eq. (\ref{eq:2.8}), $c_s >0$ in the case $V_s$ is attractive 
and $T>T_s$. 
Here, $T_s$ defined by 
$(N_F V_s)^{-1}=\ln(2\omega_{\rm D}\bar{\gamma} / \pi T_s )$ 
is the onset temperature of $s$-wave superconductivity 
in the hypothetical super-cooling state where $d_{x^2-y^2}$-wave 
superconductivity does not occur even at $T \le T_c$. 
Our study is restricted in the temperature region $T_s < T \le T_c$. 
On the other hand, $c_s <0$ in the case $V_s$ is repulsive. 

We note that, 
when the $s$-wave component $V_s$ can be neglected in the pairing 
interaction (that is, in the limit $V_s \rightarrow 0$), 
$c_s\rightarrow 0$. 
Then, the pure $d_{x^2-y^2}$-wave case is obtained. 
In the case $c_s=0$,  we recognize that $s$-wave component of 
the order parameter vanishes ($s({\bf r})$=0) from Eq. (\ref{eq:2.6}). 
Therefore, Eq. (\ref{eq:2.7}) is reduced to the one-component GL 
equation for only the $d_{x^2-y^2}$-wave component. 
There, the GL equation does not include the $s$-wave component, 
but includes the correction terms of the order $O(\gamma)$. 

In order to calculate the upper critical magnetic field $H_{c2}$ 
and the order parameter at $H_{c2}$, we consider 
the linearized GL equations. 
The annihilation and the creation operators are, respectively, 
introduced as $a= -\epsilon^{-1/2}(q_x- i q_y)$ and
$a^\dagger= -\epsilon^{-1/2}(q_x+ i q_y)$, where 
$\epsilon=4 \pi H /\phi_0=4 \pi /a_x a_y$. 
Then, 
\begin{equation}
\hat{\bf v}\cdot {\bf q}
= -\sqrt{\epsilon}(\hat{v}_+ a + \hat{v}_- a^\dagger) 
\label{eq:2.9}
\end{equation}
with $ \hat v_{\pm}=(\hat v_x \pm i \hat v_y)/2 = e^{\pm i \theta}/2 $. 
By substituting Eq. (\ref{eq:2.9}) into Eqs. (\ref{eq:2.6}) 
and (\ref{eq:2.7}), the linearized GL equations are written as
\begin{eqnarray} 
\frac{1}{\eta_0}s
&+& 
\gamma c_s \frac{2 \epsilon \xi^2}{\eta_0} 
\bigl\{ 
 \langle \phi_s^2 \hat{v}_+ \hat{v}_- \rangle [aa^\dagger]s 
+\langle \phi_s \phi_d \hat{v}_+^2 \rangle a^2 d
+\langle \phi_s \phi_d \hat{v}_-^2 \rangle a^{\dagger 2} d \bigr\} 
+ O(\gamma^2)=0, 
\label{eq:2.10}
\end{eqnarray}
\begin{eqnarray}
-\frac{1}{\eta_0}d 
&+& 
\frac{2 \epsilon \xi^2}{\eta_0}
\bigl\{ 
 \langle \phi_d^2 \hat{v}_+ \hat{v}_- \rangle [aa^\dagger]d
+\langle \phi_s \phi_d \hat{v}_+^2 \rangle a^2 s
+\langle \phi_s \phi_d \hat{v}_-^2 \rangle a^{\dagger 2} s 
\bigr\}
\nonumber \\  
&-& 
\gamma\frac{2 \epsilon^2 \xi^4}{\eta_0}
\bigl\{
 \langle \phi_d^4 \hat{v}_+^4 \rangle a^4 d 
+\langle \phi_d^4 \hat{v}_-^4 \rangle a^{\dagger 4} d
+\langle \phi_d^4 \hat{v}_+^2 \hat{v}_-^2 \rangle 
[a^2 a^{\dagger 2}] d 
\bigr\} 
+ O(\gamma^2)=0,
\label{eq:2.11}
\end{eqnarray}
where $[a^m a^{\dagger n}]$ means to take all the permutations 
of the product, such as $[ a a^\dagger ] =a a^\dagger +a^\dagger a$  
or $[ a^2 a^{\dagger 2} ]=a^2 a^{\dagger 2} +a a^\dagger a a^\dagger 
+a^\dagger a^2 a^\dagger +a a^{\dagger 2} a + a^\dagger a a^\dagger a 
+a^{\dagger 2} a^2$. 

In the investigation of the vortex lattice state, 
it is convenient to use the magnetic Bloch state,~\cite{Eilenberger} 
\begin{equation}
\psi_n({\bf r}|{\bf r}_0)
= ( \epsilon^n n! )^{-1/2} {\partial_t}^n 
\psi({\bf r}|{\bf r}_0|t) |_{t=0}, 
\label{eq:2.12}
\end{equation}
where 
\begin{eqnarray} 
\psi({\bf r}|{\bf r}_0|t)
&=&
\left(\frac{2a_y}{a_x}\right)^{1/4} 
\sum_p 
 \exp 
\Bigl[ \frac{2 \pi}{a_x a_y} 
\Bigl\{ 
-\frac{1}{2} \left( y+y_0+pa_y -2t \right)^2 
+ t^2 
-i p a_y \Bigl( x+x_0+\frac{1}{2}\zeta p a_x \Bigr)
-i y_0 x \Bigr\} \Bigr]. 
\label{eq:2.13}
\end{eqnarray}
Here, $n$ is the index of the Landau level, and there are relations 
$a \psi_n = \sqrt{n} \psi_{n-1}$ and 
$ a^\dagger \psi_n = \sqrt{n+1}\psi_{n+1}$. 
The function $\psi_0({\bf r}|{\bf r}_0)$ is the well-known 
Abrikosov solution of the vortex lattice.~\cite{Abrikosov,Eilenberger}
In Eqs. (\ref{eq:2.12}) and (\ref{eq:2.13}), ${\bf r}_0=(x_0,y_0)$ 
is related to the quasi-momentum of the Bloch state, and determines 
the position of the vortex center. 
In our presentation of figures, we set ${\bf r}_0=-({\bf r}_1+{\bf r}_2)/2$ 
so that one of the vortex centers locates at the origin of the 
coordinate. 

The order parameters can be expressed as the linear combination 
of the magnetic Bloch states as follows, 
\begin{equation}
d({\bf r})= \sum_{n=0}^\infty d_n \psi_n({\bf r}|{\bf r}_0) , \qquad 
s({\bf r})= \sum_{n=0}^\infty s_n \psi_n({\bf r}|{\bf r}_0) .
\label{eq:2.14}
\end{equation}
In the limit $\gamma \rightarrow 0$ ($T \rightarrow T_c$), 
Eqs. (\ref{eq:2.10}) and (\ref{eq:2.11}) are reduced to the 
form of the conventional GL equation.
Therefore, the order parameters are given by the Abrikosov solution 
$d({\bf r})=d_0 \psi_0({\bf r}|{\bf r}_0)$ and $s({\bf r})=0$ 
at $T \rightarrow T_c$. 
On lowering temperature below $T_c$, the corrections of the order 
$O(\gamma)$ are added to the Abrikosov solution. 
In order to investigate the corrections, the factors $d_n$ and $s_n$ 
(except for $d_0$) are expanded in powers of $\gamma$, 
\begin{eqnarray} &&
d_n= d_n^{(0)} + \gamma d_n^{(1)} + O(\gamma^2), \quad (n \ne 0) \qquad 
s_n= s_n^{(0)} + \gamma s_n^{(1)} + O(\gamma^2). 
\label{eq:2.15}
\end{eqnarray}
Now, the upper critical field is given by 
$\epsilon = 4 \pi H_{c2}/ \phi_0 $ as the function of $T$. 
We also expand $\epsilon$  in powers of $\gamma$, 
\begin{equation}
\epsilon \xi^2 = \epsilon^{(0)} \xi^2 + \gamma \epsilon^{(1)} \xi^2 
+ O(\gamma^2). 
\label{eq:2.16}
\end{equation}
Equations (\ref{eq:2.14})-(\ref{eq:2.16}) are substituted into 
Eqs. (\ref{eq:2.10}) and (\ref{eq:2.11}). 
At the order of $O(1)$, we obtain $s_n^{(0)}=0$ from 
Eq. (\ref{eq:2.10}), and 
\begin{eqnarray}
d_n^{(0)} 
&-&
 2\epsilon^{(0)} \xi^2 
\Bigl\{ 
  \langle \phi_d^2 \hat{v}_+ \hat{v}_- \rangle (2n+1)d_n^{(0)}
+ \langle \phi_s \phi_d \hat{v}_+^2 \rangle \sqrt{(n+2)(n+1)}s_{n+2}^{(0)} 
+ \langle \phi_s \phi_d \hat{v}_-^2 \rangle \sqrt{n(n-1)}s_{n-2}^{(0)} 
\Bigr\}=0 
\label{eq:2.17}
\end{eqnarray}
from Eq. (\ref{eq:2.11}).
Equation (\ref{eq:2.17}) gives 
\begin{equation}
\epsilon^{(0)} \xi^2
= \{ 2 \langle \phi_d^2 \hat{v}_+ \hat{v}_- \rangle \}^{-1} 
\label{eq:2.18}
\end{equation}
for $n=0$, and $d_n^{(0)}=0$ for $n>0$. 
On the other hand, at the order of $O(\gamma)$, we obtain 
\begin{eqnarray}
s_n^{(1)}
&+&
c_s 2\epsilon^{(0)} \xi^2  \Bigl\{ 
  \langle \phi_s \phi_d \hat{v}_+^2 \rangle \sqrt{(n+2)(n+1)}d_{n+2}^{(0)}
+ \langle \phi_s \phi_d \hat{v}_-^2 \rangle \sqrt{n(n-1)}d_{n-2}^{(0)} 
\Bigr\}=0,
\label{eq:2.19}
\end{eqnarray}
from Eq. (\ref{eq:2.10}), and 
\widetext 
\begin{eqnarray}&&
\left\{ 
1-2\epsilon^{(0)} \xi^2 \langle \phi_d^2 \hat{v}_+ \hat{v}_- \rangle 
(2n+1) \right\} d_n^{(1)} 
- 2\epsilon^{(1)}\xi^2 \langle \phi_d^2 \hat{v}_+ \hat{v}_- \rangle d_n^{(0)} 
\nonumber \\ &&
- 2\epsilon^{(0)}\xi^2  \left\{ 
 \langle \phi_s \phi_d \hat{v}_+^2 \rangle \sqrt{(n+2)(n+1)}s_{n+2}^{(1)} 
+\langle \phi_s \phi_d \hat{v}_-^2 \rangle \sqrt{n(n-1)}s_{n-2}^{(1)} \right\}
\nonumber \\ &&
+ 2(\epsilon^{(0)})^2 \xi^4 \Bigl\{ 
 \langle \phi_d^2 \hat{v}_+^4 \rangle \sqrt{(n+4)(n+3)(n+2)(n+1)}d_{n+4}^{(0)}
+\langle \phi_d^2 \hat{v}_-^4 \rangle \sqrt{n(n-1)(n-2)(n-3)}d_{n-4}^{(0)} 
\nonumber \\ &&
+ \langle \phi_d^2 \hat{v}_+^2 \hat{v}_-^2 \rangle 
3(2n^2 +2n +1)d_n^{(0)} \Bigr\}=0 
\label{eq:2.20}
\end{eqnarray}
from Eq. (\ref{eq:2.11}).
Equation (\ref{eq:2.19}) gives 
\begin{equation}
s_2^{(1)}= -\sqrt{2}c_s \frac{\langle \phi_s \phi_d \hat{v}_-^2 \rangle}
{\langle \phi_d^2 \hat{v}_+ \hat{v}_- \rangle }d_0  
\label{eq:2.21}
\end{equation}
for $n=2$, and $s_n^{(1)}=0$ for $n \ne 2$. 
From Eq. (\ref{eq:2.20}), we obtain 
\begin{equation}
\frac{\epsilon^{(1)}}{\epsilon^{(0)}}=
\frac{3 \langle \phi_d^2 \hat{v}_+^2 \hat{v}_-^2 \rangle}
     {2  \langle \phi_d^2 \hat{v}_+ \hat{v}_- \rangle^2}
+2 c_s \frac{\langle \phi_s \phi_d \hat{v}_+^2 \rangle 
\langle \phi_s \phi_d \hat{v}_-^2 \rangle}
{\langle \phi_d^2 \hat{v}_+ \hat{v}_- \rangle^2} 
\label{eq:2.22}
\end{equation}
for $n=0$, 
\begin{equation}
d_4^{(1)}= \frac{\sqrt{6}}{8}\left( 
\frac{\langle \phi_d^2 \hat{v}_-^4 \rangle}
{\langle \phi_d^2 \hat{v}_+ \hat{v}_- \rangle^2}
+2 c_s \frac{\langle \phi_s \phi_d \hat{v}_-^2 \rangle^2}
{\langle \phi_d^2 \hat{v}_+ \hat{v}_- \rangle^2} \right) d_0 
\label{eq:2.23}
\end{equation}
for $n=4$,  and $d_n^{(1)}=0$ for $n \ne 0,4$. 

As a result, $H_{c2}$ is given as 
\begin{equation}
\epsilon= \frac{4 \pi H_{c2}}{\phi_0} =
\frac{1}{2 \langle \phi_d^2 \hat{v}_+ \hat{v}_- \rangle \xi^2} 
\left\{ 1 + \gamma \frac{\epsilon^{(1)}}{\epsilon^{(0)}} + O(\gamma^2)
\right\} 
\label{eq:2.24}
\end{equation}
with the use of $\epsilon^{(1)}/\epsilon^{(0)}$ in Eq. (\ref{eq:2.22}). 
The order parameters at $H_{c2}$ are given by 
\begin{equation}
d({\bf r})=d_0 \psi_0({\bf r}|{\bf r}_0) 
+\gamma d_4^{(1)} \psi_4({\bf r}|{\bf r}_0)
+O(\gamma^2), 
\label{eq:2.25}
\end{equation}
\begin{equation}
s({\bf r})=\gamma s_2^{(1)} \psi_2({\bf r}|{\bf r}_0)
+O(\gamma^2), 
\label{eq:2.26}
\end{equation}
with the use of $d_4^{(1)}$ in Eq. (\ref{eq:2.23}) and 
$s_2^{(1)}$ in Eq. (\ref{eq:2.21}). 

Here, the $\theta$-integral is performed with adopting the forms of 
$\phi_d$ and $\phi_s$ in Eq. (\ref{eq:2.2}). 
The values of $ \langle\cdots\rangle $ appearing in this paper are 
evaluated as follows, 
\begin{eqnarray} && 
\langle \phi_d^2 \hat{v}_+ \hat{v}_- \rangle = \frac{1}{4}, \quad 
\langle \phi_d^2 \hat{v}_+^2 \hat{v}_-^2 \rangle = \frac{1}{16}, \quad 
\langle \phi_d^2 \hat{v}_-^4 \rangle = \frac{1}{32}e^{-4i \theta_0}, \quad 
\nonumber \\ &&
\langle \phi_s \phi_d \hat{v}_+^2 \rangle 
= \frac{1}{4 \sqrt{2}}e^{2i \theta_0}, \quad 
\langle \phi_s \phi_d \hat{v}_-^2 \rangle 
= \frac{1}{4 \sqrt{2}}e^{-2i \theta_0}, \quad 
\langle \phi_d^4 \rangle = \frac{3}{2}, \quad
\langle \phi_d^4 \hat{v}_+ \hat{v}_- \rangle = \frac{3}{8}. 
\label{eq:2.27}
\end{eqnarray}
Therefore, Eqs. (\ref{eq:2.24})-(\ref{eq:2.26}) are reduced to 
\begin{equation}
\epsilon= \frac{4 \pi H_{c2}}{\phi_0} =
\frac{2}{\xi^2} \left\{ 1 + \gamma \left( \frac{3}{2} +c_s \right) 
+ O(\gamma^2) \right\}, 
\label{eq:2.28}
\end{equation}
\begin{eqnarray}
d({\bf r})
&=&
d_0 \biggl\{ \psi_0({\bf r}|{\bf r}_0) 
+ \frac{\sqrt{6}}{8}\gamma \left(\frac{1}{2}+c_s \right)
e^{-4i \theta_0} \psi_4({\bf r}|{\bf r}_0) 
+O(\gamma^2) \biggr\} , 
\label{eq:2.29}
\end{eqnarray}
\begin{equation}
s({\bf r})
=d_0 \left\{ -\gamma c_s e^{-2i \theta_0} \psi_2({\bf r}|{\bf r}_0)
+O(\gamma^2) \right\}.  
\label{eq:2.30}
\end{equation}
In the correction terms of the order $O(\gamma)$ in Eqs. 
(\ref{eq:2.28})-(\ref{eq:2.30}), the terms without $c_s$ comes 
from the correction terms of the GL equation, that is, the $A_4$-terms 
in Eq. (\ref{eq:2.4}), which has been neglected so far in the 
conventional GL theory. 
The terms proportional to $c_s$ comes from the induced $s$-wave order 
parameter. 
As shown in Eq. (\ref{eq:2.30}), the effect of the induced $s$-wave 
component appears in the order $O(\gamma)$. 
It corresponds to the result $d({\bf r})/s({\bf r}) \sim 1-T/T_c$ 
suggested by the two-component GL theory.~\cite{Franz} 
From Eq. (\ref{eq:2.29}), we recognize that the correction 
of the $d_{x^2-y^2}$-wave superconductivity also appears in the same 
order of the induced $s$-wave component. 

For comparison, by exchanging $\phi_s \leftrightarrow \phi_d$ 
and putting $c_s=0$, we consider the case of an isotropic $s$-wave 
superconductor without the $s$-$d$ mixing. 
The corresponding $\theta$-integrated values are evaluated as follows, 
\begin{eqnarray} &&
\langle \phi_s^2 \hat{v}_+ \hat{v}_- \rangle = \frac{1}{4}, \quad 
\langle \phi_s^2 \hat{v}_+^2 \hat{v}_-^2 \rangle = \frac{1}{16}, \quad 
\langle \phi_s^2 \hat{v}_-^4 \rangle = 0, \quad 
\langle \phi_s^4 \rangle =1, \quad 
\langle \phi_s^4 \hat{v}_+ \hat{v}_- \rangle =\frac{1}{4}.
\label{eq:2.31}
\end{eqnarray}
In this case, from Eq. (\ref{eq:2.31}), $H_{c2}$ is written as the 
same form of Eq. (\ref{eq:2.28}) ($c_s$=0) even in an isotropic 
$s$-wave superconductor within the order $O(\gamma)$. 
On the other hand, from Eq. (\ref{eq:2.31}), 
the order parameter have the form 
$s({\bf r})=s_0 \psi_0({\bf r}|{\bf r}_0)$, that is, 
the fourth Landau level function $\psi_4({\bf r}|{\bf r}_0)$ does not 
appear in an isotropic $s$-wave case. 

Then, it is noted that the appearance of $\psi_4({\bf r}|{\bf r}_0)$ 
is the characteristic feature of the $d_{x^2-y^2}$-wave 
superconductor's vortex lattice. 
From Eq. (\ref{eq:2.29}), we see that the induced $s$-wave 
order parameter enhances the weight of $\psi_4({\bf r}|{\bf r}_0)$ 
term in the case $V_s$ is attractive ($c_s>0$). 
On the other hand, the weight of $\psi_4({\bf r}|{\bf r}_0)$ term 
is suppressed in the case $V_s$ is repulsive ($c_s<0$). 
The higher Landau level functions, 
$\psi_8({\bf r}|{\bf r}_0)$, $\psi_{12}({\bf r}|{\bf r}_0)$, $\cdots$,  
also appear in higher orders of $\gamma$. 
For example, the eighth Landau level function 
$\psi_8({\bf r}|{\bf r}_0)$ appears in the order $O(\gamma^2)$. 

As is seen from Eq. (\ref{eq:2.28}), $H_{c2}$ is enhanced by the induced 
$s$-wave order parameter in the case $V_s$ is attractive, 
which is consistent with the result of Franz {\it et al.} 
(Fig. 9 of Ref. \onlinecite{Franz}). 
On the other hand, $H_{c2}$ is suppressed in the case $V_s$ is repulsive, 
which is consistent with the result of Won and Maki 
(Fig. 1 of Ref. \onlinecite{WonMakiPRB}). 

If we set $\xi^{-2} \sim \Delta t$ with $\Delta t = 1-T/T_c$ in 
Eq. (\ref{eq:2.28}), we obtain the result corresponding to that of 
Franz {\it et al.}~\cite{Franz} 
There, the induced $s$-wave order parameter leads an upward 
curvature in the curve of $H_{c2}(T)$ near $T_c$, since the factor of 
$(\Delta t)^2$ is positive in $H_{c2}$. 
However, this upward curvature occurs even in the pure $d_{x^2-y^2}$-wave 
case ($c_s=0$) because of $\frac{3}{2}\gamma$ in Eq. (\ref{eq:2.28}), 
which is neglected in the study of Franz {\it et al.} but derived from 
the $O(\gamma)$ correction of the $d_{x^2-y^2}$-wave superconductivity. 
This upward curvature by $\frac{3}{2}\gamma$ occurs even in an isotropic 
$s$-wave superconductor in this approximation. 
These unexpected results are modified as follows by a careful consideration. 
Exactly speaking, $\xi^{-2}\sim T^2 \ln(T_c/T)$ should be expanded as 
$\xi^{-2}\sim \Delta t - \frac{3}{2} (\Delta t)^2$. 
Thus, we obtain 
\begin{equation}
\epsilon= \frac{4 \pi H_{c2}}{\phi_0} \sim 
\Delta t + \left\{c_s -\frac{3}{2}(1-\alpha) \right\} (\Delta t)^2 .
\label{eq:2.32}
\end{equation}
Now, in the pure $d_{x^2-y^2}$-wave case and the isotropic 
$s$-wave case ($c_s=0$), the factor of $(\Delta t)^2 $ is negative 
and the curve of $H_{c2}(T)$ shows a downward curvature near $T_c$. 
For $c_s > \frac{3}{2}(1-\alpha)$, the factor of $(\Delta t)^2$ 
changes to positive and the curve of $H_{c2}(T)$ shows an upward 
curvature. 

As is seen from Eq. (\ref{eq:2.21}), the amplitude of the induced 
$s$-wave order parameter is determined by not only the parameter of 
the gradient coupling between $d_{x^2-y^2}$- and $s$-wave order 
parameters, $\langle \phi_s \phi_d \hat{v}_-^2 \rangle$, but also 
the parameter of the pairing interaction, $c_s$ defined 
in Eq. (\ref{eq:2.8}). 
The amplitude of the induced $s$-wave order parameter is proportional 
to the product of both parameters. 

From Eq. (\ref{eq:2.26}), the spatial variation of the induced $s$-wave 
order parameter has the same structure as that of the second Landau level 
function $\psi_2({\bf r}|{\bf r}_0)$. 
Figure \ref{fig:2} is the case of a square lattice ($a_y/a_x=0.5$, 
$\zeta=0.5$). 
In Fig. \ref{fig:2} (a), we show the amplitude of the 
induced $s$-wave order parameter, 
$|s({\bf r})/\gamma s_2^{(1)}|=|\psi_2({\bf r}|{\bf r}_0)|$. 
The amplitude is suppressed near the vortex of the induced $s$-wave 
order parameter. 
To show it clearly, we schematically present the position of the 
vortices and the winding number of their phase in Fig. \ref{fig:2} (b). 
At the vortex center of the $d_{x^2-y^2}$-wave superconductivity 
(its winding number is +1), the induced $s$-wave order parameter 
has a vortex with the winding number $-1$. 
It is consistent with the result of the single vortex 
case.\cite{Berlinsky,Franz,Ren,Xu,Soininen,IchiokaS}
Further, the $s$-wave order parameter has extra vortices with 
the winding number +2, which locates at the farthest points 
from the vortices of the $d_{x^2-y^2}$-wave superconductivity. 
In total, we obtain the winding number +1 for the phase of the 
$s$-wave order parameter when we go round about the boundary of 
a unit cell in the vortex lattice. 
The amplitude and the winding number of the $s$-wave order parameter 
are not affected by the orientation $\theta_0$ of the vortex lattice. 
Figure \ref{fig:3} is the case of an oblique lattice with $73^\circ$ 
($a_y/a_x=0.676$, $\zeta=0.5$). 
In Fig. \ref{fig:3} (a), we show the amplitude of the induced 
$s$-wave order parameter, which is consistent with the result 
of Berlinsky {\it et al.} (Fig. 3 (b) in Ref. \onlinecite{Berlinsky}. 
The $a$ axis and the $b$ axis are exchanged 
each other in our calculation.)  
The corresponding position of the vortices and their winding 
number are shown in Fig. \ref{fig:3} (b). 
In the oblique lattice case, the vortex with the winding number +2 in the 
square lattice case splits into two vortices with the winding 
number +1. 
In the limit of a triangular lattice, these vortices with the winding 
number +1 locates at the farthest points from the vortices of 
the $d_{x^2-y^2}$-wave superconductivity.

The spatial variation of the $d_{x^2-y^2}$-wave order parameter 
in Eq. (\ref{eq:2.29}) is presented later in Sec. \ref{sec:5}.

\section{Current and magnetic field}
\label{sec:3}

We calculate the current and its induced magnetic field 
around vortices near $H_{c2}$. 
Magnetic field ${\bf H}({\bf r})$ is divided into an external 
field ${\bf H}_0=(0,0,H_0)$ and an internal field 
${\bf h}({\bf r})$ which is induced by the current ${\bf j}({\bf r})$,
\begin{equation} 
{\bf H}({\bf r})= {\bf H}_0 +{\bf h}({\bf r}). 
\label{eq:3.1}
\end{equation}
Amplitude of the order parameter is small near $H_{c2}$. 
Therefore, as for the current and magnetic field, we consider 
terms up to the order O($|d_0|^2)$. 
The expression for the current density is derived as follows 
from the Gor'kov equation with including the correction terms 
of the order $O(\gamma)$  
(As for the detail of the derivation, see Appendix A of 
Ref. \onlinecite{Enomoto}.),  
\begin{eqnarray}
{\bf j}({\bf r}) 
&\equiv&
\xi \nabla \times {\bf H}({\bf r})/(\phi_0/\xi^2)
=\xi \nabla \times {\bf h}({\bf r})/(\phi_0/\xi^2)
\label{eq:3.2} \\ 
&=&-\frac{2\pi}{\kappa^2 \eta_0^2} \Bigl[  \xi 
\langle \hat{\bf v} \{(\hat{\bf v}\cdot{\bf q})\Delta({\bf r},\theta) \}
\Delta^\ast({\bf r},\theta) \rangle 
\nonumber \\ &&
- \gamma \xi^3 \Bigl\{ 
\langle \hat{\bf v} \{ (\hat{\bf v}\cdot{\bf q})^3 \Delta({\bf r},\theta) \}
\Delta^\ast({\bf r},\theta) \rangle
+\langle \hat{\bf v} \{ (\hat{\bf v}\cdot{\bf q})^2 \Delta({\bf r},\theta) \}
\{ (\hat{\bf v}\cdot{\bf q})\Delta({\bf r},\theta) \}^\ast \rangle 
\Bigr\} 
+ O(\gamma^2) \Bigr] +{\rm c.c.}, 
\label{eq:3.3} 
\end{eqnarray}
where terms with $|\Delta|^4$ (that is, terms in the  order $O(|d_0|^4)$) 
are neglected. 
In Eq. (\ref{eq:3.3}), 
\begin{equation}
\kappa^{-2}=\frac{4 |e| N_F v_F^2 \beta \xi^2 \eta_0^2}{3c \phi_0} 
= \frac{3}{8 \pi^2}\left(\frac{T_c}{T}\right)^2 \kappa_{\rm GL}^{-2}, 
\label{eq:3.4} 
\end{equation}
where the GL parameter $\kappa_{\rm GL}$ is given by 
\begin{equation}
\kappa_{\rm GL}=\frac{9}{7 \pi \zeta(3) N_F}
\left(\frac{\hbar c}{|e|}\right)^2 
\frac{( \pi k_B T_c )^2}{(\hbar v_F)^4}.
\label{eq:3.5}
\end{equation}

With the substitution of Eqs. (\ref{eq:2.1}), (\ref{eq:2.9}) 
and (\ref{eq:2.14}) into Eq. (\ref{eq:3.3}), the current density 
is written as 
\begin{equation}
{\bf j}({\bf r})=\frac{2 \pi \sqrt{\epsilon \xi^2}}
{\kappa^2 \eta_0^2 }\sum_{n_1,n_2} ({\bf j})_{n_1,n_2} 
\psi_{n_1}({\bf r}|{\bf r}_0) \psi_{n_2}^\ast({\bf r}|{\bf r}_0) , \qquad 
\label{eq:3.6}
\end{equation}
where 
\widetext  
\begin{eqnarray}
({\bf j})_{n_1,n_2}
&=& 
\langle \phi_d^2 \hat{\bf v}\hat{v}_+ \rangle 
\left(
 \sqrt{n_1+1}d_{n_1+1}d_{n_2}^\ast + \sqrt{n_2}d_{n_1}d_{n_2-1}^\ast
\right) 
+\langle \phi_d^2 \hat{\bf v}\hat{v}_- \rangle
\left(
 \sqrt{n_1}d_{n_1-1}d_{n_2}^\ast + \sqrt{n_2+1}d_{n_1}d_{n_2+1}^\ast 
\right) 
\nonumber \\ &&
+\langle \phi_s^2 \hat{\bf v}\hat{v}_+ \rangle
\left(
 \sqrt{n_1+1}s_{n_1+1}s_{n_2}^\ast + \sqrt{n_2}s_{n_1}s_{n_2-1}^\ast 
\right) 
+\langle \phi_s^2 \hat{\bf v}\hat{v}_- \rangle
\left(
 \sqrt{n_1}s_{n_1-1}s_{n_2}^\ast + \sqrt{n_2+1}s_{n_1}s_{n_2+1}^\ast 
\right)
\nonumber \\ && 
+\langle \phi_s \phi_d \hat{\bf v}\hat{v}_+ \rangle 
\left( 
  \sqrt{n_1+1}s_{n_1+1}d_{n_2}^\ast + \sqrt{n_2}s_{n_1}d_{n_2-1}^\ast 
 +\sqrt{n_1+1}d_{n_1+1}s_{n_2}^\ast + \sqrt{n_2}d_{n_1}s_{n_2-1}^\ast 
\right)
\nonumber \\ &&
+\langle \phi_s \phi_d \hat{\bf v}\hat{v}_- \rangle 
\left(
  \sqrt{n_1}s_{n_1-1}d_{n_2}^\ast + \sqrt{n_2+1}s_{n_1}d_{n_2+1}^\ast 
 +\sqrt{n_1}d_{n_1-1}s_{n_2}^\ast + \sqrt{n_2+1}d_{n_1}s_{n_2+1}^\ast 
\right)
\nonumber \\ 
&-&
\gamma\epsilon \xi^2 \Bigl\{
 \langle \phi_d^2 \hat{\bf v}\hat{v}_+^3 \rangle
\Bigl( 
  \sqrt{(n_1+3)(n_1+2)(n_1+1)} d_{n_1+3} d_{n_2}^\ast 
 +\sqrt{n_2 (n_2-1) (n_2-2)} d_{n_1} d_{n_2-3}^\ast
\nonumber \\ &&
  +\sqrt{(n_1+2)(n_1+1)n_2} d_{n_1+2} d_{n_2-1}^\ast 
  +\sqrt{(n_1+1)n_2(n_2-1)} d_{n_1+1} d_{n_2-2}^\ast 
\Bigr)
\nonumber \\ &&
+\langle \phi_d^2 \hat{\bf v}\hat{v}_-^3 \rangle
\Bigl( 
  \sqrt{n_1(n_1-1)(n_1-2)} d_{n_1-3} d_{n_2}^\ast 
 +\sqrt{(n_2+3)(n_2+2)(n_2+1)} d_{n_1} d_{n_2+3}^\ast
\nonumber \\ &&
 +\sqrt{n_1 (n_1-1)(n_2+1)} d_{n_1-2} d_{n_2+1}^\ast 
 +\sqrt{n_1 (n_2+2)(n_2+1)} d_{n_1-1} d_{n_2+2}^\ast
\Bigr)
\nonumber \\ &&
+\langle \phi_d^2 \hat{\bf v}\hat{v}_+^2 \hat{v}_- \rangle
\Bigl(
  (3n_1+2n_2+4)\sqrt{n_1+1} d_{n_1+1} d_{n_2}^\ast 
 +(2n_1+3n_2+1)\sqrt{n_2} d_{n_1} d_{n_2-1}^\ast 
\nonumber \\ &&
 +\sqrt{(n_1+2)(n_1+1)(n_2+1)} d_{n_1+2} d_{n_2+1}^\ast 
 +\sqrt{n_1 n_2 (n_2-1) } d_{n_1-1} d_{n_2-2}^\ast 
\Bigr)
\nonumber \\ &&
+\langle \phi_d^2 \hat{\bf v}\hat{v}_+ \hat{v}_-^2 \rangle
\Bigl(
  (3n_1+2n_2+1)\sqrt{n_1} d_{n_1-1} d_{n_2}^\ast 
 +(2n_1+3n_2+4)\sqrt{n_2+1} d_{n_1} d_{n_2+1}^\ast 
\nonumber \\ &&
 +\sqrt{n_1 (n_1-1) n_2} d_{n_1-2} d_{n_2-1}^\ast 
 +\sqrt{(n_1+1)(n_2+2)(n_2+1)} d_{n_1+1} d_{n_2+2}^\ast 
\Bigr) \Bigr\}
+O(\gamma^2). 
\label{eq:3.7}
\end{eqnarray}
Among the correction terms of the order $O(\gamma)$ in Eq. (\ref{eq:3.7}), 
we neglect the terms including $s_n$, since $s_n$ is in the 
order $O(\gamma)$ as shown in Eq. (\ref{eq:2.26}).
Equation (\ref{eq:3.7}) satisfies a relation 
\begin{equation} 
({\bf j})_{n_2,n_1}=\{({\bf j})_{n_1,n_2} \}^\ast  , 
\label{eq:3.8}
\end{equation}
which ensures that ${\bf j}({\bf r})$ is real. 
By substituting the results of $\epsilon$ and the  order parameters  
$d_n$ and $s_n$ in Sec \ref{sec:2} [Eqs. (\ref{eq:2.15}), 
(\ref{eq:2.16}), (\ref{eq:2.18}) and (\ref{eq:2.21})-(\ref{eq:2.23})] 
into Eq. (\ref{eq:3.7}) and by introducing 
\begin{equation}
j_\pm = (j_x \pm i j_y)/2, 
\label{eq:3.9}
\end{equation}
we obtain 
\widetext  
\begin{eqnarray}
(j_+)_{1,0}
&=& 
\langle \phi_d^2  \hat{v}_+ \hat{v}_- \rangle |d_0|^2 
+\gamma \left(
 \langle \phi_s \phi_d \hat{v}_+^2 \rangle \sqrt{2} s_2^{(1)} d_0^\ast 
 - \epsilon^{(0)}\xi^2 
 \langle \phi_d^2  \hat{v}_+^2 \hat{v}_-^2 \rangle 4 |d_0|^2  \right)
\nonumber \\ 
&=&
 \langle \phi_d^2  \hat{v}_+ \hat{v}_- \rangle \left\{ 1
-2 \gamma \left( 
\frac{\langle \phi_d^2  \hat{v}_+^2 \hat{v}_-^2 \rangle}
{\langle \phi_d^2  \hat{v}_+ \hat{v}_- \rangle^2}
+c_s \frac{\langle \phi_s \phi_d \hat{v}_+^2 \rangle 
\langle \phi_s \phi_d \hat{v}_-^2 \rangle}
{\langle \phi_d^2  \hat{v}_+ \hat{v}_- \rangle^2} \right) + O(\gamma^2) 
\right\} |d_0|^2 , 
\label{eq:3.10}
\end{eqnarray}
\begin{eqnarray}
(j_-)_{3,0}
&=&
 \gamma \left( 
\langle \phi_d^2  \hat{v}_+ \hat{v}_- \rangle 2 d_4^{(1)} d_0^\ast 
+\langle \phi_s \phi_d \hat{v}_-^2 \rangle \sqrt{3} s_2^{(1)} d_0^\ast 
-\epsilon^{(0)}\xi^2 \langle \phi_d^2  \hat{v}_-^4 \rangle \sqrt{6}|d_0|^2 
\right)
\nonumber \\ 
&=&
 \langle \phi_d^2  \hat{v}_+ \hat{v}_- \rangle \left\{ 
-\frac{\sqrt{6}}{4}\gamma \left( 
\frac{\langle \phi_d^2  \hat{v}_-^4 \rangle}
{\langle \phi_d^2  \hat{v}_+ \hat{v}_- \rangle^2} 
+2 c_s \frac{\langle \phi_s \phi_d \hat{v}_-^2 \rangle^2}
{\langle \phi_d^2  \hat{v}_+ \hat{v}_- \rangle^2} 
\right) + O(\gamma^2)
\right\} |d_0|^2 , 
\label{eq:3.11}
\end{eqnarray}
\begin{eqnarray} 
(j_+)_{5,0}
&=&
 \gamma 
\langle \phi_d^2  \hat{v}_+ \hat{v}_- \rangle \sqrt{5} d_4^{(1)} d_0^\ast 
\nonumber \\ 
&=&
 \langle \phi_d^2  \hat{v}_+ \hat{v}_- \rangle \left\{
\frac{\sqrt{30}}{8}\gamma \left(
\frac{\langle \phi_d^2  \hat{v}_-^4 \rangle}
{\langle \phi_d^2  \hat{v}_+ \hat{v}_- \rangle^2}
+2 c_s \frac{\langle \phi_s \phi_d \hat{v}_-^2 \rangle^2}
{\langle \phi_d^2  \hat{v}_+ \hat{v}_- \rangle^2}
\right) + O(\gamma^2)
\right\} |d_0|^2 , 
\label{eq:3.12}
\end{eqnarray}
\begin{eqnarray}
(j_+)_{2,1}
&=&
 \gamma \left(
\langle \phi_s \phi_d \hat{v}_+^2 \rangle s_2^{(1)} d_0^\ast 
-\epsilon^{(0)}\xi^2 \langle \phi_d^2  \hat{v}_+^2 \hat{v}_-^2 \rangle
\sqrt{2} |d_0|^2 \right)
\nonumber \\ 
&=& \langle \phi_d^2  \hat{v}_+ \hat{v}_- \rangle \left\{
-\frac{\sqrt{2}}{2}\gamma \left(
\frac{\langle \phi_d^2  \hat{v}_+^2 \hat{v}_-^2 \rangle}
{\langle \phi_d^2  \hat{v}_+ \hat{v}_- \rangle^2}
+2 c_s \frac{\langle \phi_s \phi_d \hat{v}_+^2 \rangle 
\langle \phi_s \phi_d \hat{v}_-^2 \rangle}
{\langle \phi_d^2  \hat{v}_+ \hat{v}_- \rangle^2}
\right) + O(\gamma^2)
\right\} |d_0|^2 , 
\label{eq:3.13}
\end{eqnarray}
\begin{eqnarray}
(j_-)_{4,1}
&=&
\gamma
\langle \phi_d^2  \hat{v}_+ \hat{v}_- \rangle \sqrt{5} d_4^{(1)} d_0^\ast
\nonumber \\ 
&=&
 \langle \phi_d^2  \hat{v}_+ \hat{v}_- \rangle \left\{
\frac{\sqrt{6}}{8}\gamma \left(
\frac{\langle \phi_d^2  \hat{v}_-^4 \rangle}
{\langle \phi_d^2  \hat{v}_+ \hat{v}_- \rangle^2}
+2 c_s \frac{\langle \phi_s \phi_d \hat{v}_-^2 \rangle^2}
{\langle \phi_d^2  \hat{v}_+ \hat{v}_- \rangle^2}
\right) + O(\gamma^2)
\right\} |d_0|^2 . 
\label{eq:3.14}
\end{eqnarray}
The other components $(j_\pm)_{n_1,n_2}$ for $n_1 \ge n_2$ are zero within 
the order $O(\gamma)$. 
As a result, from Eqs. (\ref{eq:3.8}) and (\ref{eq:3.9}), 
the current density is given as follows by using 
Eqs. (\ref{eq:3.10}) - (\ref{eq:3.14}), 
\begin{eqnarray} 
j_x=2 {\rm Re} \{ 
&&  
(j_+)_{1,0} \psi_1 \psi_0^\ast 
 +(j_-)_{3,0} \psi_3 \psi_0^\ast  +(j_+)_{5,0} \psi_5 \psi_0^\ast
 +(j_+)_{2,1} \psi_2 \psi_1^\ast  +(j_-)_{4,1} \psi_4 \psi_1^\ast \}, 
\label{eq:3.15} 
\end{eqnarray}
\begin{eqnarray}
j_y=2 {\rm Im} \{  
&&
-(j_+)_{1,0} \psi_1 \psi_0^\ast
 +(j_-)_{3,0} \psi_3 \psi_0^\ast  -(j_+)_{5,0} \psi_5 \psi_0^\ast
 -(j_+)_{2,1} \psi_2 \psi_1^\ast  +(j_-)_{4,1} \psi_4 \psi_1^\ast \}. 
\label{eq:3.16}
\end{eqnarray}

The magnetic field induced by the current is obtained from the 
Maxwell equation. 
When we neglect terms of the order $O(|d_0|^4)$, 
the induced magnetic field ${\bf h}({\bf r})$ has the following form, 
\begin{equation}
{\bf h}({\bf r})= \sum_{n_1,n_2} ({\bf h})_{n_1,n_2} 
\psi_{n_1}({\bf r}|{\bf r}_0) \psi_{n_2}^\ast({\bf r}|{\bf r}_0) , 
\label{eq:3.17}
\end{equation}
where $({\bf h})_{n_1,n_2}$ satisfies a relation 
\begin{equation}
({\bf h})_{n_2,n_1}=\{({\bf h})_{n_1,n_2} \}^\ast  
\label{eq:3.18}
\end{equation}
for ${\bf h}({\bf r})$ to be real. 
In our case, ${\bf h}({\bf r})$ has only the $z$-component, 
${\bf h}({\bf r})=(0,0,h_z({\bf r}))$. 

From the Maxwell equation (\ref{eq:3.2}), we have relations 
\begin{eqnarray} 
(j_+)_{n_1,n_2}=\frac{\sqrt{\epsilon \xi^2}}{2}\frac{\xi^2}{\phi_0} \Bigl\{ 
&& 
-\sqrt{n_1}(h_z)_{n_1-1,n_2}
+\sqrt{n_2+1}(h_z)_{n_1,n_2+1} \Bigr\}, 
\label{eq:3.19} 
\end{eqnarray}
\begin{eqnarray}
(j_-)_{n_1,n_2}=\frac{\sqrt{\epsilon \xi^2}}{2}\frac{\xi^2}{\phi_0} \Bigl\{ 
&&
\sqrt{n_1+1}(h_z)_{n_1+1,n_2}
-\sqrt{n_2}(h_z)_{n_1,n_2-1} \Bigr\} . 
\label{eq:3.20}
\end{eqnarray}
We also have a relation 
\begin{eqnarray} && 
\sqrt{n_1+1}(j_+)_{n_1+1,n_2}+\sqrt{n_1}(j_-)_{n_1-1,n_2}
-\sqrt{n_2+1}(j_-)_{n_1,n_2+1}-\sqrt{n_2}(j_+)_{n_1,n_2-1}=0
\label{eq:3.21}
\end{eqnarray}
from ${\rm div}{\bf j}=0$. 
Therefore, $(h_z)_{n_1,n_2}$ is written as follows from 
Eqs. (\ref{eq:3.10})-(\ref{eq:3.14}) and (\ref{eq:3.19})-(\ref{eq:3.21}), 
\begin{eqnarray}
(h_z)_{0,0}
&=&
-\frac{2}{\sqrt{\epsilon \xi^2}}\frac{\phi_0}{\xi^2} 
\sum_{n=1}^\infty \frac{1}{\sqrt{n}} (j_+)_{n,n-1}
=
h_z^{(0)}+ \gamma (h_z^{(1)})_{0,0} + O(\gamma^2), 
\label{eq:3.22}
\end{eqnarray}
\begin{eqnarray}
(h_z)_{1,1}
&=&
-\frac{2}{\sqrt{\epsilon \xi^2}}\frac{\phi_0}{\xi^2} 
\sum_{n=2}^\infty \frac{1}{\sqrt{n}} (j_+)_{n,n-1}
=
\gamma (h_z^{(1)})_{1,1} + O(\gamma^2), 
\label{eq:3.23}
\end{eqnarray}
\begin{eqnarray}
(h_z)_{4,0}
&=&
-\frac{2}{\sqrt{\epsilon \xi^2}}\frac{\phi_0}{\xi^2}
\left\{ -\frac{1}{2}(j_-)_{3,0} \right\}
=
\gamma (h_z^{(1)})_{4,0} + O(\gamma^2),
\label{eq:3.24}
\end{eqnarray}
where 
\begin{equation}
h_z^{(0)}=-\frac{4 \pi \langle \phi_d^2  \hat{v}_+ \hat{v}_- \rangle }
{\kappa^2 \eta_0^2}|d_0|^2  , \quad (<0)
\label{eq:3.25}
\end{equation}
\begin{equation}
\frac{(h_z^{(1)})_{0,0}}{h_z^{(0)}}=- \left(
\frac{5 \langle \phi_d^2  \hat{v}_+^2 \hat{v}_-^2 \rangle}
{2 \langle \phi_d^2  \hat{v}_+ \hat{v}_- \rangle^2} 
+ 3 c_s \frac{ \langle \phi_s \phi_d  \hat{v}_+^2  \rangle 
\langle \phi_s \phi_d  \hat{v}_-^2  \rangle}
{\langle \phi_d^2  \hat{v}_+ \hat{v}_- \rangle^2} \right) , 
\label{eq:3.26}
\end{equation}
\begin{equation}
\frac{(h_z^{(1)})_{1,1}}{h_z^{(0)}}=  - \left(
\frac{ \langle \phi_d^2  \hat{v}_+^2 \hat{v}_-^2 \rangle}
{2 \langle \phi_d^2  \hat{v}_+ \hat{v}_- \rangle^2} 
+ c_s \frac{ \langle \phi_s \phi_d  \hat{v}_+^2  \rangle 
\langle \phi_s \phi_d  \hat{v}_-^2  \rangle}
{\langle \phi_d^2  \hat{v}_+ \hat{v}_- \rangle^2} \right) , 
\label{eq:3.27}
\end{equation}
\begin{equation}
\frac{(h_z^{(1)})_{4,0}}{h_z^{(0)}} = \frac{\sqrt{6}}{8} \left( 
\frac{\langle \phi_d^2  \hat{v}_-^4 \rangle}
{\langle \phi_d^2  \hat{v}_+ \hat{v}_- \rangle^2} 
+2 c_s \frac{\langle \phi_s \phi_d \hat{v}_-^2 \rangle^2}
{\langle \phi_d^2  \hat{v}_+ \hat{v}_- \rangle^2}     \right) . 
\label{eq:3.28}
\end{equation}
The other components $(h_z)_{n_1,n_2}$ for $n_1 \ge n_2$ 
vanish within the order $O(\gamma)$. 
As a result, the induced magnetic field is written as 
\begin{eqnarray}
h_z({\bf r})
&=&
  \left\{ h_z^{(0)} + \gamma (h_z^{(1)})_{0,0} \right\} |\psi_0|^2 
+ \gamma (h_z^{(1)})_{1,1} |\psi_1|^2 
+ 2 \gamma {\rm Re}\left\{ (h_z^{(1)})_{4,0} \psi_4 \psi_0^\ast \right\}
+ O(\gamma^2). 
\label{eq:3.29}
\end{eqnarray}

By using the $\theta$-integrated values in Eq. (\ref{eq:2.27}), 
$h_z({\bf r})$ in Eq. (\ref{eq:3.29}) is reduced to 
\begin{eqnarray}
h_z({\bf r})
&=&
h_z^{(0)} \biggl[ \left\{ 
1- \frac{5}{2}\gamma \left( 1+\frac{3}{5}c_s \right) \right\} |\psi_0|^2    
-\frac{1}{2} \gamma ( 1+ c_s ) |\psi_1|^2 
+\frac{\sqrt{6}}{8}\gamma \left( 1+ 2 c_s \right) {\rm Re}
\left\{ e^{-4i \theta_0} \psi_4 \psi_0^\ast \right\} + O(\gamma^2) \biggr]. 
\label{eq:3.30}
\end{eqnarray}
The spatial variation of the magnetic field in Eq. (\ref{eq:3.30}) 
is presented later in Sec. \ref{sec:5}. 

When we consider the case of an isotropic $s$-wave superconductor 
without the $s$-$d$ mixing ($\phi_s \leftrightarrow \phi_d$, $c_s=0$), 
by using Eqs. (\ref{eq:2.31}) we obtain 
\begin{eqnarray}
h_z({\bf r})
&=&
h_z^{(0)} \biggl\{ 
\left( 1- \frac{5}{2}\gamma  \right) |\psi_0|^2
-\frac{1}{2} \gamma |\psi_1|^2
+ O(\gamma^2) \biggr\}, 
\label{eq:3.31}
\end{eqnarray}
where $(h_z^{(1)})_{4,0}$ vanishes. 
From the comparison between Eqs. (\ref{eq:3.30}) and (\ref{eq:3.31}), 
it is concluded that the appearance of the $\psi_4 \psi_0^\ast$ term 
in the magnetic field distribution is the characteristic feature of 
the $d_{x^2-y^2}$-wave superconductor's vortex lattice. 
As is seen from Eq. (\ref{eq:3.30}), the contribution of the 
$\psi_4 \psi_0^\ast$ term is enhanced by the induced $s$-wave order 
parameter in the case $V_s$ is attractive ($c_s>0$). 
On the other hand, its contribution is suppressed in the case $V_s$
is repulsive ($c_s<0$).

\section{Orientation and unit cell shape of the vortex lattice}
\label{sec:4}

To know the equilibrium state of the vortex lattice, we search 
the free energy minimum about the orientation and the unit cell shape 
of the vortex lattice in the vicinity of $H_{c2}$. 
The free energy $F$ is calculated by following the method of 
Abrikosov~\cite{Abrikosov} and Takanaka.~\cite{TakanakaPTP} 
As a result, the free energy of the vortex lattice is written 
as follows (the detailed derivation is given in Appendix \ref{sec:a}),   
\begin{equation}
\frac{F}{F_0}=\kappa^2 \frac{{\bf B}^2}{(\phi_0/\xi^2)^2}
-\frac{\displaystyle \left\{ \kappa^2 
\frac{({\bf B} - {\bf H}_{c2})\cdot \overline{{\bf h}}}
{(\phi_0/\xi^2)^2 |d_0|^2}
\right\}^2 }
{\displaystyle \kappa^2 \frac{\overline{{\bf h}}^2 - \overline{{\bf h}^2}}
{(\phi_0/\xi^2)^2 |d_0|^4}
+ \frac{D_4[d_c,s_c]}{2 \eta_0^4 |d_0|^4} } , 
\label{eq:4.1}
\end{equation} 
with $F_0 = N_F \xi^2 \eta_0^2 \ln(T_c/T)$ and the magnetic flux density  
\begin{equation}
{\bf B}= \overline{\bf{H}({\bf r})}={\bf H_0}+ 
\overline{\bf{h}({\bf r})} . 
\label{eq:4.2}
\end{equation}
Here, we write $\int(\cdots)d{\bf r}/\xi^2 = \overline{(\cdots)}$. 
In Eq. (\ref{eq:4.1}), $D_4[d_c,s_c]$ is given by 
\begin{eqnarray}
\frac{D_4[d_c,s_c]}{|d_0|^4} 
&=&\frac{2}{3} \langle \phi_d^4 \rangle \frac{\overline{|d|^4}}{|d_0|^4} 
- \gamma 
\frac{\langle \phi_d^4 \hat{v}_+ \hat{v}_- \rangle}
{3 \langle \phi_d^2 \hat{v}_+ \hat{v}_- \rangle}
\left( 
5 \overline{|\psi_0|^4} -2 \overline{|\psi_1|^2 |\psi_0|^2} 
\right) 
+ O(\gamma^2) 
\label{eq:4.3}
\end{eqnarray}
with
\begin{equation}
\frac{\overline{|d|^4}}{|d_0|^4} = \overline{|\psi_0|^4} 
+4\gamma {\rm Re} \left\{ \frac{d_4^{(1)}}{d_0}
\overline{\psi_4 \psi_0^\ast |\psi_0|^2 } \right\} 
+ O(\gamma^2). 
\label{eq:4.4}
\end{equation}
When a magnetic field has only the $z$ component as in the present case, 
Eq. (\ref{eq:4.1}) is reduced to 
\begin{eqnarray}
\frac{F}{F_0}
&=&
\kappa^2 \frac{B^2}{(\phi_0/\xi^2)^2}
- \frac{(B-H_{c2})^2}{(\phi_0/\xi^2)^2}
\left\{ 1+ \frac{2\bar{\kappa}^2 C(a_y/a_x,\zeta,\theta_0) }
{(\overline{h_z} / h_z^{(0)} )^2}   
 \right\}^{-1}, 
\label{eq:4.5}
\end{eqnarray}
where we use the notations  
\begin{equation}
C(a_y/a_x,\zeta,\theta_0) 
\equiv 
\frac{D_4[d_c,s_c]}{|d_0|^4}
- \frac{1}{2\bar{\kappa}^2}\frac{\overline{h_z^2}}{(h_z^{(0)})^2} , 
\label{eq:4.6}
\end{equation}
\begin{eqnarray}
\bar{\kappa}
&\equiv&
\phi_0|d_0|^2 \left( 2 \kappa h_z^{(0)} \xi^2 \eta_0^2  \right)^{-1} 
=
\left( 2\sqrt{6} \langle \phi_d^2 \hat{v}_+ \hat{v}_- \rangle \right)^{-1} 
\frac{T}{T_c} \kappa_{\rm GL}. 
\label{eq:4.7}
\end{eqnarray}
By using Eq. (\ref{eq:3.29}) and the orthogonalized condition 
$\overline{\psi_{n_1} \psi_{n_2}^\ast}=\delta_{n_1,n_2}$, 
the quantities $\overline{h_z}$ in Eq. (\ref{eq:4.5}) and 
$\overline{h_z^2}$ in Eq. (\ref{eq:4.6}) are given as follows, 
\begin{equation}
\frac{\overline{h_z}}{h_z^{(0)}}= 1+ \gamma \left\{ 
\frac{(h_z^{(1)})_{0,0}}{h_z^{(0)}} + \frac{(h_z^{(1)})_{1,1}}{h_z^{(0)}} 
\right\} + O(\gamma^2),
\label{eq:4.8}
\end{equation}
\begin{eqnarray}
\frac{\overline{h_z^2}}{(h_z^{(0)})^2}
&=&
\left\{ 1+ 2 \gamma 
\frac{(h_z^{(1)})_{0,0}}{h_z^{(0)}} \right\} \overline{|\psi_0|^4}
+ 2 \gamma \frac{(h_z^{(1)})_{1,1}}
{h_z^{(0)}} \overline{|\psi_0|^2 |\psi_1|^2}
+ 4 \gamma {\rm Re}\left\{ \frac{(h_z^{(1)})_{4,0}}{h_z^{(0)}} 
\overline{\psi_4 \psi_0^\ast |\psi_0|^2 } \right\} 
+ O(\gamma^2). 
\label{eq:4.9}
\end{eqnarray}
In Eq. (\ref{eq:4.5}),  
$ 2\bar{\kappa}^2 ( \overline{h_z}/ h_z^{(0)} )^{-2} 
C(a_y/a_x,\zeta,\theta_0) $ corresponds 
to $(2 \kappa^2 -1 )\beta_A $ of the Abrikosov's theory.~\cite{Abrikosov} 
In the limit $\gamma \rightarrow 0$, Eq. (\ref{eq:4.5}) is reduced to 
his result. 

The quantities $\overline{|\psi_0|^4}$, 
$\overline{\psi_4 \psi_0^\ast |\psi_0|^2 }$ and 
$\overline{|\psi_1|^2 |\psi_0|^2}$ are calculated 
as follows from Eq. (\ref{eq:2.12}),  
\begin{equation}
\overline{|\psi_0|^4}
= \left(\frac{a_y}{a_x}\right)^{1/2}\sum_{n,m}\exp\left\{ 
-\pi \frac{a_y}{a_x}(m^2+n^2)\right\} \cos(2 \pi \zeta m n),
\label{eq:4.10}
\end{equation}
\begin{eqnarray} 
\overline{\psi_4 \psi_0^\ast |\psi_0|^2 }
&=&
 \frac{1}{2\sqrt{6}}\left(\frac{a_y}{a_x}\right)^{1/2} \sum_{n,m}
\biggl\{ \frac{3}{4} 
-3(m-n)^2 \pi \frac{a_y}{a_x} 
+ (m-n)^4 \left( \pi \frac{a_y}{a_x} \right)^2 \biggr\} 
\nonumber \\ && \times 
\exp\left\{-\pi \frac{a_y}{a_x}(m^2+n^2) + i 2 \pi \zeta m n \right\}, 
\label{eq:4.11}
\end{eqnarray}
\begin{equation}
\overline{|\psi_1|^2 |\psi_0|^2}=\frac{1}{2} \overline{|\psi_0|^4} . 
\label{eq:4.12}
\end{equation}
These quantities depend on the parameters $a_y/a_x$ and $\zeta$ 
characterizing the unit cell shape of the vortex lattice. 
Therefore, the free energy (\ref{eq:4.5}) depends on the unit cell shape 
of the vortex lattice through the quantities 
$\overline{|\psi_0|^4}$, $\overline{\psi_4 \psi_0^\ast |\psi_0|^2 }$ 
and $\overline{|\psi_1|^2 |\psi_0|^2}$.
For the triangular lattice case ($a_y/a_x= \sqrt{3}/2$, $\zeta=0.5$), 
$\overline{\psi_4 \psi_0^\ast |\psi_0|^2 }=0$.
On the other hand, the orientation $\theta_0$ of the vortex lattice 
affects the free energy (\ref{eq:4.5}) through $d_4^{(1)}$ in 
Eq. (\ref{eq:2.23}) and $(h_z^{(1)})_{4,0}$ in Eq. (\ref{eq:3.28})
as is seen from Eq. (\ref{eq:2.27}). 
Since $\overline{h_z}$ does not depend on the orientation and the 
unit cell shape, the free energy  (\ref{eq:4.5}) has its minimum 
at the vortex lattice structure where $C(a_y/a_x,\zeta,\theta_0)$ 
defined in Eq. (\ref{eq:4.6}) takes a minimum value. 

From Eqs. (\ref{eq:4.3}), (\ref{eq:4.9}) and (\ref{eq:4.12}), 
$C(a_y/a_x,\zeta,\theta_0)$ is written as 
\widetext 
\begin{eqnarray}
C(a_y/a_x,\zeta,\theta_0)
&=&
\left[ \frac{2}{3} \langle \phi_d^4 \rangle 
      -\gamma\frac{4 \langle \phi_d^4 \hat{v}_+ \hat{v}_- \rangle}
                  {3 \langle \phi_d^2 \hat{v}_+ \hat{v}_- \rangle} 
-\frac{1}{2\bar{\kappa}^2} \left\{ 1- \gamma \left(
    \frac{11 \langle \phi_d^2 \hat{v}_+^2 \hat{v}_-^2 \rangle}
         {2 \langle \phi_d^2 \hat{v}_+ \hat{v}_- \rangle^2 } 
+7c_s \frac{\langle \phi_s \phi_d \hat{v}_+^2 \rangle 
            \langle \phi_s \phi_d \hat{v}_-^2 \rangle } 
           {\langle \phi_d^2 \hat{v}_+ \hat{v}_- \rangle^2} \right) 
\right\} \right] \overline{|\psi_0|^4}
\nonumber \\ && 
+\frac{\sqrt{6}}{2}\left(  \frac{2}{3} \langle \phi_d^4 \rangle 
 - \frac{1}{2 \bar{\kappa}^2} \right) \gamma 
{\rm Re}\left\{\left( 
    \frac{ \langle \phi_d^2 \hat{v}_-^4 \rangle }
         {\langle \phi_d^2 \hat{v}_+ \hat{v}_- \rangle^2}
 +2c_s \frac{\langle \phi_s \phi_d \hat{v}_-^2 \rangle^2}
            {\langle \phi_d^2 \hat{v}_+ \hat{v}_- \rangle^2} \right)
\overline{\psi_4 \psi_0^\ast |\psi_0|^2 } \right\} + O(\gamma^2)  .
\label{eq:4.13}
\end{eqnarray}
When we use the $\theta$-integrated values in Eq. (\ref{eq:2.27}),  
Eq. (\ref{eq:4.13}) is reduced to 
\begin{eqnarray}
C
(a_y/a_x,\zeta,\theta_0) 
&=&\left[ 1-2\gamma  -\frac{1}{2 \bar{\kappa}^2}
\left\{ 1-\frac{1}{2} \gamma (11 + 7 c_s ) \right\} \right]
\overline{|\psi_0|^4}
\nonumber \\ &&
+\frac{\sqrt{6}}{4}\left(1 - \frac{1}{2 \bar{\kappa}^2} \right)
\gamma(1+2c_s){\rm Re}\left\{ e^{-4i\theta_0} 
\overline{\psi_4 \psi_0^\ast |\psi_0|^2 } \right\} 
+ O(\gamma^2)  . 
\label{eq:4.14}
\end{eqnarray}

The orientation of the vortex lattice is easily determined from 
Eq. (\ref{eq:4.14}). 
With respect to $\theta_0$, the minimum of $C(a_y/a_x,\zeta,\theta_0)$ 
occurs at 
\begin{equation}
\theta_0 = \frac{1}{4} \left\{ {\rm arg}
\left( \overline{\psi_4 \psi_0^\ast |\psi_0|^2 } \right) - \pi \right\} , 
\label{eq:4.15} 
\end{equation}
for $c_s > -0.5$, where 
${\rm arg}( \overline{\psi_4 \psi_0^\ast |\psi_0|^2 } )$ means 
the phase of the complex function $\overline{\psi_4 \psi_0^\ast |\psi_0|^2}$. 
Here, $\bar{\kappa} \gg 1$ since we consider the case of an 
extreme type II superconductor. 
For $c_s < -0.5$, where $V_s$ is strong repulsive, the minimum occurs at 
\begin{equation}
\theta_0 = \frac{1}{4} {\rm arg}
\left( \overline{\psi_4 \psi_0^\ast |\psi_0|^2 } \right) . 
\label{eq:4.16}
\end{equation}
Then, the orientation $\theta_0$ is rotated by $\pi/4$ from that of the 
case $c_s > -0.5$. 
For both cases of Eqs. (\ref{eq:4.15}) and (\ref{eq:4.16}), 
the minimum of $C(a_y/a_x,\zeta,\theta_0)$ as a function of $\theta_0$ 
is given by 
\begin{eqnarray}
&\bar{C}&(a_y/a_x,\zeta) 
\equiv 
{\rm min}_{\theta_0}\{ C(a_y/a_x,\zeta,\theta_0) \} 
\nonumber \\ 
&=& 
\left[ 1-2\gamma -\frac{1}{2 \bar{\kappa}^2}
\left\{ 1-\frac{1}{2} \gamma (11 + 7 c_s ) \right\} \right]
\overline{|\psi_0|^4}
-\frac{\sqrt{6}}{4}\left(1 - \frac{1}{2 \bar{\kappa}^2} \right)
\gamma |1+2c_s| |\overline{\psi_4 \psi_0^\ast |\psi_0|^2 }| 
+ O(\gamma^2) . 
\label{eq:4.17}
\end{eqnarray}

We numerically estimate the minimum of $\bar{C}(a_y/a_x,\zeta)$ 
with respect to $a_y/a_x$ and  $\zeta$. 
The term of $\overline{|\psi_0|^4}$ prefers a triangular lattice, 
which is the well-known result in the conventional GL theory.~\cite{Abrikosov} 
Then, at $T=T_c$ (i.e., $\gamma=0$), triangular lattice is realized as 
the equilibrium vortex lattice. 
On the other hand, the term of 
$-|\overline{\psi_4 \psi_0^\ast |\psi_0|^2 }|$ prefers a square 
lattice.
On lowering temperature from $T_c$ (i.e., increasing $\gamma$) 
along the curve of $H_{c2}(T)$,  
the contribution of $-|\overline{\psi_4 \psi_0^\ast |\psi_0|^2 }|$ 
term increases. 
Then, the vortex lattice is deformed from a triangular lattice 
to a square lattice.
There are three processes for this deformation, 
which are enumerated as (a), (b) and (c) in the following. 

(a) On lowering temperature from $T_c$, the ratio $a_y/a_x$ 
gradually varies from $\sqrt{3}/2=0.866$ (triangular lattice) to 
0.5 (square lattice) with preserving $\zeta=0.5$. 
It means that the vortices form a shape of an isosceles triangle 
with $BO=BA$ as schematically presented in Fig. \ref{fig:4}. 
Since the function $\overline{\psi_4 \psi_0^\ast |\psi_0|^2 }$  
in Eq. (\ref{eq:4.11}) is real and negative in this case, 
the orientation is given by $\theta_0=0$ for $c_s > -0.5$. 
(In the following, our calculation is restricted in the case $c_s > -0.5$.) 
Therefore, the base $OA$ of the isosceles triangle is along the $a$ axis 
direction. 

(b) On lowering temperature from $T_c$, the parameters $(a_y/a_x,\zeta)$ 
gradually vary from $(\sqrt{3}/2, 0.5)$ [triangular lattice] 
to $(1,0)$ [square lattice] 
with keeping the shape of an isosceles triangle with $OA=OB$. 
In this process, $\theta_0$ varies from $30^\circ$ to $45^\circ$ 
so that the base $AB$ of the isosceles triangle is along the 
$b$ direction. 

(c) On lowering temperature from $T_c$, the parameters $(a_y/a_x,\zeta)$
gradually vary from $(\sqrt{3}/2, 0.5)$ to $(1,1)$ [square lattice] 
with keeping the shape of an isosceles triangle with $AB=AO$. 
In this process, $\theta_0$ varies from $-30^\circ$ to $-45^\circ$ 
so that the base $BO$ of an isosceles triangle is along the 
$b$ direction. 

The cases (b) and (c) are related each other by the relation 
$C(a_y/a_x,1-\zeta,-\theta_0)=C(a_y/a_x,\zeta,\theta_0)$. 
As a result, the cases (a), (b) and (c) are equivalent and give the same 
orientation and unit cell shape. 
The vortex lattice is deformed from a triangular lattice to a square 
lattice with keeping the shape of an isosceles triangle 
on lowering temperature from $T_c$. 
As for the orientation of the vortex lattice, the base of the isosceles 
triangle is along the $a$ axis or the $b$ axis direction. 
Therefore, at the lower temperature region, the equilibrium state of the 
vortex lattice is a square lattice tilted by $\pi/4$ from the $a$ axis, 
which is consistent with the result of Won and Maki~\cite{WonMakiPRB}  

To see the temperature dependence of the unit cell shape, 
we consider the case (a) in the limit $\bar{\kappa} \gg 1$. 
In this limit, Eq. (\ref{eq:4.17}) is reduced to  
\begin{eqnarray}
&\bar{C}&(a_y/a_x,\zeta) 
= (1-2\gamma ) \overline{|\psi_0|^4}
-\frac{\sqrt{6}}{4}
\gamma |1+2c_s| |\overline{\psi_4 \psi_0^\ast |\psi_0|^2 }| 
+ O(\gamma^2) ,  
\label{eq:4.18}
\end{eqnarray}
where it is enough to consider only the contribution of 
the condensation energy. 
The ratio $a_y/a_x$ in the equilibrium vortex lattice  structure 
is evaluated from the minimum of $\bar{C}(a_y/a_x,\zeta=0.5) $
in Eq. (\ref{eq:4.18}). 
In Fig. \ref{fig:5}, we present the temperature dependence of 
the ratio $a_y/a_x$ along the curve of $H_{c2}(T)$ 
for various mixing of the $s$-wave component, 
$c_s=-0.2$, $-0.1$, 0, 0.2, 0.4. 
As shown in Fig. \ref{fig:5}, the deformation from a triangular 
lattice starts just below $T_c$ on lowering temperature. 
In the pure $d_{x^2-y^2}$-wave case ($c_s=0$), the square lattice 
is realized for $T \le 0.858 T_c$. 
In the case $V_s$ is attractive ($c_s>0$), the square lattice is 
realized from higher temperatures 
($T \le 0.888 T_c$ for $c_s= 0.2$, $T \le 0.908 T_c$ for $c_s= 0.4$), 
since the contribution of 
$-|\overline{\psi_4 \psi_0^\ast |\psi_0|^2 }|$
in Eq. (\ref{eq:4.18}) is enhanced due to the induced $s$-wave 
order parameter. 
On the other hand, in the case $V_s$ is repulsive ($c_s<0$), 
the temperature region of the square lattice is suppressed 
($T \le 0.837 T_c$ for $c_s= -0.1$, $T \le 0.807 T_c$ for $c_s= -0.2$). 

When we consider the case of an isotropic $s$-wave superconductor 
without the $s$-$d$ mixing 
($\phi_s \leftrightarrow \phi_d$, $c_s=0$ in Eq. (\ref{eq:4.13})), 
by using Eq. (\ref{eq:2.31}) we obtain 
\begin{eqnarray}
&C&(a_y/a_x,\zeta,\theta_0) 
= \left\{ \frac{2}{3}(1-2\gamma ) 
-\frac{1}{2 \bar{\kappa}^2}\left( 1-\frac{11}{2}\gamma \right) \right\}
\overline{|\psi_0|^4}
+ O(\gamma^2) . 
\nonumber \\  &&
\label{eq:4.19}
\end{eqnarray}
In this isotropic $s$-wave case, the contribution of 
$\overline{\psi_4 \psi_0^\ast |\psi_0|^2 }$ is absent from 
$C(a_y/a_x,\zeta,\theta_0)$.
Then, the minimum of $\overline{|\psi_0|^4}$ determines the 
equilibrium vortex lattice structure. 
A triangular lattice is, therefore, realized in all the temperature 
region. 
It is consistent with the well-known result in the conventional GL  
theory.~\cite{Abrikosov} 

In the calculation of Won and Maki,~\cite{WonMakiPRB,WonMaki} 
they obtain the equilibrium state by estimating the Abrikosov parameter 
$\beta_A= \langle \overline{|\Delta({\bf r},\theta)|^4} \rangle/
(\langle \overline{|\Delta({\bf r},\theta)|^2}\rangle )^2$. 
It seems to be an analogy of the conventional GL theory. 
Within the order $O(\gamma)$, we obtain 
\begin{eqnarray}
\beta_A 
&=&
\overline{|d|^4}/|d_0|^4 + O(\gamma^2)
=
 \overline{|\psi_0|^4}-\frac{\sqrt{6}}{4}
\gamma (1+2c_s) {\rm Re}\{e^{-4i\theta_0} 
\overline{\psi_4 \psi_0^\ast |\psi_0|^2 } \}
+ O(\gamma^2) .
\label{eq:4.20}
\end{eqnarray}
where it is noted that $1-2\gamma$ in the first term of 
Eq. (\ref{eq:4.14}) is changed to 1.
However, $\beta_A$ itself should be modified by the correction of the 
order $O(\gamma)$. 
In our calculation, we use the function $C(a_y/a_x,\zeta,\theta_0)$ in 
Eq. (\ref{eq:4.14}) instead of $\beta_A$.  
It changes the result quantitatively. 
Dominant terms neglected in the estimate by $\beta_A$ come from the 
gradient term of the $\Delta^3$-order such as 
\{$\langle \phi_d^4(\hat{\bf v}\cdot\hat{\bf q})^2 \rangle d\} |d|^2 $ 
in the GL equation. 
If we consider the correction of the non-local term 
$\langle \phi_d^2 (\hat{\bf v}\cdot\hat{\bf q})^4 \rangle d$ 
in the calculation of the $\Delta$-linear order, 
in the estimate of the free energy 
we have to include also the gradient terms of the $\Delta^3$-order 
as correction, which appear within the same order of $\gamma$. 

While Won and Maki considered only the cases of a triangular lattice 
and a square lattice, they obtain the result that the square lattice is stabilized at $T < 0.88 T_c$ in the pure $d_{x^2-y^2}$-case. 
This temperature is not so different from ours. 
In their result, as repulsion of the $s$-wave component increases in the 
interaction, the temperature region of the square lattice is suppressed 
(Fig. 4 of Ref. \onlinecite{WonMakiPRB}). 
This tendency of the repulsive case is consistent with our result. 
Won and Maki~\cite{WonMakiPRB,WonMaki} suggested that the experimental 
result of an oblique lattice~\cite{Keimer,Maggio,Renner} can be explained 
by the square lattice structure obtained here and by the scaling of 
the coordinate system due to the difference of the coherence lengths 
between the $a$ axis and the $b$ axis.

\section{Spatial variation of the vortex lattice structure
in the pure $d_{x^2-y^2}$-wave case}
\label{sec:5}

As clarified in previous sections, the characteristic features of 
the $d_{x^2-y^2}$-wave superconductor's vortex lattice is produced 
by the appearance of the fourth Landau level function 
$\psi_4({\bf r}|{\bf r}_0)$ in the vortex lattice structure. 
Due to this effect, the vortex lattice is deformed from a triangular 
lattice to a square lattice as shown in Sec. \ref{sec:4}. 
In this section, we consider the effect of $\psi_4({\bf r}|{\bf r}_0)$ 
on the spatial variation of the vortex lattice structure in a 
$d_{x^2-y^2}$-wave superconductor by using the results of 
Secs. \ref{sec:2} and \ref{sec:3}. 
Here, we consider the case of the pure $d_{x^2-y^2}$-wave superconductor 
($c_s=0$). 
Then, in the following characteristic features of the vortex lattice, 
there is no contribution of the induced $s$-wave order parameter. 

First, we consider the spatial variation of the $d_{x^2-y^2}$-wave 
order parameter, which is given by Eq. (\ref{eq:2.29}). 
The contour plot of the amplitude, $|d({\bf r})/d_0|$,  is presented 
in Fig. \ref{fig:6}, where the core region of the vortex is darkly shaded. 
We consider the case of a square lattice at a lower temperature 
[Figs. \ref{fig:6} (a) and (b)] and the case of an oblique lattice 
at an intermediate temperature [Fig. \ref{fig:6} (c)]. 
The parameter $\gamma$ is chosen by following the result of 
Fig. \ref{fig:5}. 

As for the case of a square lattice ($a_y/a_x$=0.5, $\zeta$=0.5), 
we use $\gamma$=0.2 ($T/T_c$=0.802). 
In addition to the case of an equilibrium orientation 
$\theta_0=0^\circ$ in Fig. \ref{fig:6} (a), we also show the case of 
an unstable orientation $\theta_0=45^\circ$ in Fig. \ref{fig:6} (b) 
to clarify the contribution of $\psi_4({\bf r}|{\bf r}_0)$. 
In both figures, the $a$ axis and the $b$ axis are along the 
horizontal and vertical directions.  
(In Fig. \ref{fig:6} (a), the $x$ axis and the $y$ axis are also 
along the horizontal and vertical directions. 
In Fig. \ref{fig:6} (b), the $x$ axis and the $y$ axis are rotated by 
$45^\circ$ from the case of Fig. \ref{fig:6} (a).)
Due to the property of a $d_{x^2-y^2}$-wave superconductor, 
the amplitude of the order parameter is slightly suppressed along 
the direction of the $a$ axis and the $b$ axis around a vortex, 
which has been exhibited in the calculation  of the single vortex 
case.~\cite{IchiokaF,Enomoto} 
On the other hand, due to the property of the vortex lattice, 
the amplitude of the order parameter is slightly suppressed 
along the directions of nearest-neighbor vortices, 
which has been exhibited in the calculation of the isotropic $s$-wave 
superconductor case.~\cite{IchiokaLDOS}   
Therefore, when the $a$ axis and the $b$ axis agree with the 
nearest neighbor directions as in the case of Fig. \ref{fig:6} (b), 
the suppression along the nearest neighbor directions is enhanced 
and the fourfold symmetric vortex core structure is emphasized. 
On the other hand, when the $a$ axis and the $b$ axis are away 
from the nearest-neighbor directions as in the case of 
Fig. \ref{fig:6} (a), the suppression of both effects cancels each other. 
Therefore, the vortex core structure is reduced to a cylindrically 
symmetric one. 
It seems that the gain of the condensation energy in the case of 
Fig. \ref{fig:6} (a) is more efficient than that of Fig. \ref{fig:6} (b). 
Then, the equilibrium vortex lattice structure prefers the 
configuration of Fig. \ref{fig:6} (a).  

Figure \ref{fig:6} (c) is the case of an oblique lattice with $73^\circ$ 
($a_y/a_x$=0.676, $\zeta$=0.5) for the equilibrium orientation $\theta_0=0$, 
where we use $\gamma$=0.115 ($T/T_c$=0.881). 
In this figure, the $a$ axis and the $b$ axis 
(and the $x$ axis and the $y$ axis) 
are along the horizontal and vertical directions.
Also in this oblique lattice case for $\theta_0=0$,  
the vortex core has cylindrically symmetric structure 
since the $a$ axis and the $b$ axis 
are away from the nearest-neighbor directions. 

Next, we consider the spatial variation of the current and 
the induced magnetic field around vortices in each case of 
Fig. \ref{fig:6}. 
The current is calculated from Eqs. (\ref{eq:3.15}) and (\ref{eq:3.16}), 
where we use the $\theta$-integrated values of Eq. (\ref{eq:2.27}). 
Figure \ref{fig:7} shows the contour plot for the amplitude of the current, 
$|{\bf j}({\bf r})|/(\pi \sqrt{\epsilon \xi^2}/2\kappa^2 \eta^2)$. 
The corresponding magnetic field is calculated from Eq. (\ref{eq:3.30}). 
Figure \ref{fig:8} shows the contour plot of the magnetic field, 
$h_z({\bf r})/|h_z^{(0)}|$. 
The contour lines are also interpreted as the stream lines of the 
current. 

The calculation of the single vortex lattice case shows that 
$|{\bf j}({\bf r})|$ and $h_z({\bf r})$ are fourfold symmetric 
around a vortex core in a $d_{x^2-y^2}$-wave 
superconductor.~\cite{IchiokaF,Enomoto} 
There, $|{\bf j}({\bf r})|$ has four peaks around the vortex core 
at the direction of $45^\circ$ from the $a$ axis or the $b$ axis. 
Reflecting this current distribution, the magnetic field extends 
along the $a$ axis and the $b$ axis directions around the vortex core. 
This is because the field is strongly screened by the induced current 
along the $45^\circ$ directions. 
These fourfold symmetric structures of $|{\bf j}({\bf r})|$ and 
$h_z({\bf r})$ are also recognized in our calculation of the vortex 
lattice case, which becomes clear with increasing $\gamma$ 
(lowering temperature). 

In the square lattice case at a lower temperature ($\gamma=0.2$), 
we compare the case of the equilibrium orientation $\theta_0=0^\circ$ 
in Figs. \ref{fig:7} (a) and \ref{fig:8} (a) with that of the unstable 
orientation $\theta_0=45^\circ$ in Figs. \ref{fig:7} (b) and \ref{fig:8} (b).
For the equilibrium case $\theta_0=0^\circ$, the screening current 
flows in the narrow region around each vortex compared with the 
unstable case $\theta_0=45^\circ$. 
Then, for $\theta_0=0^\circ$, the enhancement of $h_z({\bf r})$ 
at each vortex core localizes in the narrow region, 
and $h_z({\bf r})$ is reduced to be a flat distribution at the 
region apart from vortex cores. 

Figure \ref{fig:7} (c) and \ref{fig:8} (c) are the oblique lattice 
case at the intermediate temperature ($\gamma=0.115$). 
There, fourfold symmetric structure of $|{\bf j}({\bf r})|$ and 
$h_z({\bf r})$ are not seen so clearly due to the small $\gamma$ 
and also due to the broken fourfold symmetry of the vortex lattice structure. 

Another way to investigate the spatial variation of $h_z({\bf r})$  
is to consider the magnetic field distribution function defined as 
\begin{equation}
P(h)= \int_{\rm unit \ cell}\delta(h-h_z({\bf r}))d{\bf r}/a_x a_y , 
\label{eq:5.1}
\end{equation}
which is related to the resonance line shape in the $\mu$SR and 
NMR experiments. 
In Fig. \ref{fig:9}, we show $P(h)$ as a function of $h/|h_z^{(0)}|$ 
for the equilibrium state of the vortex lattice at various 
temperatures; $\gamma=$0 (a), 0.06 (b), 0.115 (c), 0.15 (d) and 
0.2 (e), which correspond to $T/T_c=$1, 0.936, 0.881, 0.848 
and 0.802, respectively. 
The case of a pure $d_{x^2-y^2}$-wave superconductor given by 
Eq. (\ref{eq:3.30}) is presented by solid lines. 
The parameters of the vortex lattice are chosen by following the 
result of Fig. \ref{fig:5}. 
Then, $\zeta=0.5$, $\theta_0=$0 and 
$a_y/a_x=$0.866 (a), 0.799 (b), 0.676 (c), 0.5 (d), 0.5 (e). 
In the case of a triangular lattice [line (a)] and a square lattice 
[lines (d) and (e)], $P(h)$ has single peak structure. 
On the other hand, in the case of an oblique lattice [lines (b) and (c)], 
$P(h)$ has double peaks structure. 
These structures are consistent with the result of 
Franz {\it et al.},\cite{Franz,Affleck}  
while the oblique lattice is derived from 
the induced $s$-wave order parameter in their theory.
As explained by them, the double peak structure occurs 
in an oblique lattice because the saddle points of $h_z({\bf r})$ 
between two vortices are not equivalent in the ${\bf r}_1$ direction 
and in the ${\bf r}_2$ direction. 

For comparison, we also show $P(h)$ of the isotropic $s$-wave case 
given by Eq. (\ref{eq:3.31}), 
which is presented by dotted lines in Fig. \ref{fig:9}. 
In this $s$-wave case, the equilibrium state is a triangular lattice 
($\zeta=0.5$, $a_y/a_x=0.866$).
We compare the case of a square lattice in a pure $d_{x^2-y^2}$-wave 
superconductor with that of a triangular lattice in a isotropic $s$-wave 
superconductor. 
At $\gamma=0.16$ [line (d)], the peak of the square lattice case locates 
at the right side of the peak of the triangular lattice case, 
which is consistent with the result of the 
conventional GL theory.~\cite{Fetter} 
In the $d_{x^2-y^2}$-wave case, with increasing $\gamma$ (decreasing $T$), 
the lower edge of $P(h)$ approaches the peak position 
and the peak height increases as is seen from the line (e).   
This change of the distribution reflects the result of Fig. \ref{fig:8} (a), 
where $h_z({\bf r})$ has uniform distribution outside the core region. 
In the isotropic $s$-wave case, on the other hand, 
the distribution of $P(h)$ does not so vary on lowering temperature.

Some of the above-mentioned features are similar to that of Berlinsky 
{\it et al}.,~\cite{Berlinsky,Franz,Affleck} where the characteristic 
features of the vortex lattice are derived 
from the contribution of the induced $s$-wave order parameter. 
However, it is important to notice that the characteristic features suggested 
by our calculation occur even in the pure $d_{x^2-y^2}$-wave case 
where the induced $s$-wave order parameter is absent. 
Its origin is the $A_4$-terms in Eq. (\ref{eq:2.4}), which is in the 
order $O(\ln(T_c/T))$ in the dimensionless form of the GL equation 
and neglected in the conventional GL equation. 
The contribution of these correction terms increases with lowering 
temperature from $T_c$ and leads to the fourfold symmetric vortex core 
structure.  
It also induces the $\psi_4({\bf r}|{\bf r}_0)$ term in the vortex 
lattice structure and leads to the above-mentioned characteristic 
features of the vortex lattice in a $d_{x^2-y^2}$-wave superconductor.

If the induced $s$-wave order parameter cannot be neglected, 
the contribution of the $\psi_4({\bf r}|{\bf r}_0)$ term increases 
for $c_s>0$ and decreases for $c_s<0$, as shown in the result of 
Secs. \ref{sec:2} and \ref{sec:3}. 
Therefore, the above-mentioned characteristic features,  
which are produced by the $\psi_4({\bf r}|{\bf r}_0)$ term,  
are  enhanced in the case $V_s$ is attractive ($c_s>0$) and 
suppressed in the case $V_s$ is repulsive ($c_s<0$).

\section{Summary and discussions}
\label{sec:6}

We investigate the vortex lattice structure in a $d_{x^2-y^2}$-wave 
superconductor  near $H_{c2}$ in the framework of the extended GL theory. 
We determine the orientation and the unit cell shape of the vortex lattice. 
As for the unit cell shape, vortices form a shape of an isosceles triangle 
as shown in Fig. \ref{fig:4}. 
The vertical angle gradually varies from $60^\circ$ to $90^\circ$ when 
temperature is lowered from $T_c$. 
Then, a square lattice is realized at the low temperature region 
$T < 0.86 T_c$ for the pure $d_{x^2-y^2}$-wave case ($c_s=0$). 
As for the orientation, the base of an isosceles triangle is 
along the $a$ axis (or the $b$ axis). 

We also calculate the spatial variation of the vortex structure, 
such as the $d_{x^2-y^2}$-wave order parameter, the current and 
the induced magnetic field. 
As temperature is enough lowered, the fourfold symmetric core structure 
clearly appears in our calculation of the vortex lattice case. 
This structure is consistent with that of the single vortex case. 
The amplitude of the current has four peaks around each vortex in the 
$45^\circ$ direction from the $a$ axis. 
Then, the induced magnetic field extends along the $a$ axis and 
the $b$ axis direction. 
As for the $d_{x^2-y^2}$-wave order parameter, the amplitude around the 
vortex is slightly suppressed along the $a$ axis and the $b$ axis 
direction due to the nature of the $d_{x^2-y^2}$-wave superconductivity. 
On the other hand, the amplitude is slightly suppressed in the direction 
of the nearest-neighbor vortex. 
When the $a$ axis and the $b$ axis are away from the nearest-neighbor 
directions (i.e., the case of the orientation $\theta_0=0^\circ$), 
the suppression of both effects cancels each other. 
Therefore, the vortex core structure is reduced to a cylindrically 
symmetric one.
This structure seems to most effectively gain the condensation energy. 
It is the reason why the equilibrium state prefers
 the orientation $\theta_0=0^\circ$. 

Some of the above-mentioned characteristic features 
of the vortex lattice are similar to that of 
Berlinsky {\it et al.},~\cite{Berlinsky,Franz,Affleck} 
where the origin of the characteristic features in their result 
is the mixing of the induced $s$-wave order parameter. 
However, it is noted that the characteristic features 
obtained in this paper appear even in the pure $d_{x^2-y^2}$-wave case 
where the induced $s$-wave order parameter is absent. 
The origin of these characteristic features is the higher order terms 
of the order parameter and the higher derivatives (or non-local terms) 
which are neglected in the conventional GL theory. 
These correction terms derived from the Gor'kov equation are higher order 
in the small parameter $\ln(T_c/T)$. 
Therefore, their contribution vanishes at $T_c$, but becomes 
increasingly important  upon lowering temperature from $T_c$. 
Some of these correction terms induce the contribution of the fourth 
Landau level function $\psi_4({\bf r}|{\bf r}_0)$ to the vortex lattice 
structure within the order $O(\ln(T_c/T))$. 
This contribution of $\psi_4({\bf r}|{\bf r}_0)$ leads to the 
characteristic features of the vortex lattice in the $d_{x^2-y^2}$-wave  
superconductor, such as the deformation from a triangular lattice and 
the fourfold symmetric structure around a vortex core.

We also estimate the effect of the induced $s$-wave order parameter. 
The amplitude of the $s$-wave order parameter depends on $V_s$, 
the $s$-wave component of the pairing interaction, 
through the parameter $c_s$ defined in Eq. (\ref{eq:2.8}) in our model. 
When the contribution of the induced $s$-wave order parameter 
cannot be neglected, the factor of the $\psi_4({\bf r}|{\bf r}_0)$ 
term increases (decreases) for $c_s>0$ ($c_s<0$) in the expression of 
the vortex lattice structure. 
Then, the characteristic features, 
which is produced by the $\psi_4({\bf r}|{\bf r}_0)$ term,  
are enhanced (suppressed) in the case $V_s$ is attractive (repulsive). 

In the following, we give some comments.

At further low temperatures, we have to consider further higher order 
derivatives in the GL equation. 
Then, the contribution of higher Landau level 
functions $\psi_8({\bf r}|{\bf r}_0)$, $\psi_{12}({\bf r}|{\bf r}_0)$, 
$\cdots$, appear in the vortex lattice structure, 
which are in the order $O([\ln(T_c/T)]^n)$ ($n \ge 2$). 
The investigation of these higher order correction is now in progress. 

The fourfold symmetric behavior is also observed when magnetic field is 
applied in the direction parallel to the $ab$ plane. 
When the direction of the field is rotated within the plane, 
the $H_{c2}$ (Refs. \onlinecite{Hanaguri} and \onlinecite{Koike}) 
and the torque (Ref. \onlinecite{Ishida}) experiments show the fourfold 
symmetric behavior. 
They are explained by the contribution of the higher order derivative 
terms (or non-local terms) in the GL equation as in this 
paper.~\cite{WonMaki,Takanaka,TakanakaJPSJ} 
Also in this parallel field case, if the induced $s$-wave oder parameter 
exists, the fourfold symmetric behavior is enhanced (suppressed)  
when $V_s$ is attractive (repulsive).~\cite{Sugiyama} 

The amplitude of the induced $s$-wave order parameter depends on $V_s$, 
the $s$-wave component of the interaction, through the parameter 
$c_s$ in Eq. (\ref{eq:2.8}). 
Then, next, we want to know the value of $V_s$ appropriate for 
the high-$T_c$ materials. 
However, it seems to be a difficult problem, which is deeply related 
to the pairing mechanism of the high-$T_c$ superconductors. 
If the $s$-wave order parameter is defined as the on-site pairing, 
$V_s$ may be strongly repulsive since the high-$T_c$ materials are 
strongly correlated electron systems. 
However, $s$-wave order parameter is defined as the extended $s$-wave, 
we cannot exclude the possibility of the attractive $V_s$. 
The extended $s$-wave is derived from the non-local pairing 
between nearest neighbor atomic site, which is the same origin of the 
$d_{x^2-y^2}$-wave pairing. 

We note that it is not clear whether the two-component GL theory 
can be applied to the case of strongly repulsive $V_s$. 
In the limit of strong repulsive $V_s$ ($c_s \rightarrow -\infty$), 
the induced $s$-wave order parameter completely destroys the 
$d_{x^2-y^2}$-wave superconductivity, 
which is seen from Eq. (\ref{eq:2.32}) for example. 
In this strong repulsive case, we have to reconsider from the outset 
whether we can introduce the $s$-wave order parameter or not. 

So far, we consider the case of the isotropic Fermi surface. 
If the Fermi surface is anisotropic, 
the average $\langle \cdots \rangle$ is modified to 
$\langle \cdots \rho(\theta) \rangle$, where $\rho(\theta)$ is the 
$\theta$-dependent factor of the density of states on the Fermi surface 
and $\int\rho(\theta) d \theta / 2\pi =1$. 
And $|\hat{\bf v}|$ has the $\theta$-dependence. 

The origin of the fourfold symmetric core structure and the 
deformation from a triangular lattice is the appearance of 
$\psi_4({\bf r}|{\bf r}_0)$ in the vortex lattice structure. 
The contribution of $\psi_4({\bf r}|{\bf r}_0)$ is proportional to 
the quantity $\langle \phi_d^2 v_-^4 \rho \rangle$ in the pure 
$d_{x^2-y^2}$-wave case. 
Even in the $s$-wave superconductor ($\phi_d \rightarrow \phi_s$), 
when the structure of the Fermi surface or the energy 
gap ($|\phi_s|$) have the 
anisotropic component with fourfold symmetry, the contribution of 
$\psi_4({\bf r}|{\bf r}_0)$ appears since 
$\langle \phi_s^2 v_-^4 \rho \rangle \ne 0$. 
Then, the fourfold symmetric core structure and the deformation 
from a triangular lattice occur at lower temperature and 
higher magnetic field. 
The square lattice observed by De Wilde {\it et al.}\cite{DeWilde} 
in the STM image of the vortex lattice on ${\rm LuNi_2B_2C}$ 
seems to be this case. 
And Eskildsen {\it et al.}\cite{Eskildsen} reported the deformation from a  
triangular lattice to a square lattice with increasing magnetic field 
by the SANS and magnetic decoration observation of the flux line lattice 
on ${\rm ErNi_2B_2C}$, where the orientation and the unit cell shape of the
vortex lattice in the deforming process agree with our results. 
In the $d_{x^2-y^2}$-wave superconductor, 
$|\langle \phi_d^2 v_-^4 \rho \rangle|$ takes a large value since 
$\phi_d^2$ has fourfold symmetric node structure.  
In the $s$-wave superconductor, to introduce a large value of 
$|\langle \phi_s^2 v_-^4 \rho \rangle|$ comparable to the 
$d_{x^2-y^2}$-wave case, $\phi_s^2 |v_-|^4 \rho $ should have a large 
anisotropy with fourfold symmetry such as the form of 
$(1+\cos 4\theta)/2$.

\section*{Acknowledgments}
We are grateful to K. Takanaka for informing us the detail 
of his calculation. 
We also would like to thank N. Hayashi and T. Sugiyama for 
fruitful discussions. 
One of the authors (M. I.) would like to thank 
the Japan Society for the Promotion of Science for financial support.

\appendix
\section{Free energy of the vortex lattice state} 
\label{sec:a}

The expression of the free energy $F$ is determined so that 
$\partial F / \partial d^\ast({\bf r})=0$ and 
$\partial F / \partial s^\ast({\bf r})=0$, respectively, give 
the GL equations (\ref{eq:2.6}) and (\ref{eq:2.7}). 
Thus, the free energy is written as 
\begin{eqnarray}
\frac{F}{F_0}
&=& 
\kappa^2 \frac{\overline{{\bf H}({\bf r})^2}}{(\phi_0 /\xi^2)^2}
+\eta_0^{-2} \biggl\{ - \overline{|d|^2} 
+ (c_s \gamma)^{-1}\overline{|s|^2}  
+ \overline{d^\ast K d } + \overline{s^\ast K_{ss} s} 
+ \overline{s^\ast K_{sd} d} + \overline{d^\ast K_{sd} s}  \biggr\} 
\nonumber \\ 
&+&
 \frac{1}{2}\eta_0^{-4}D_4[d,s]
+ \frac{1}{3} \eta_0^{-6} \overline{d^\ast R_5} + O(\gamma^2) 
\label{eq:a.1}
\end{eqnarray}
with 
\begin{eqnarray}
D_4[d,s]
&=&
 \overline{d^\ast R_3}
+ \frac{2}{3} \langle \phi_s^4 \rangle \overline{|s|^4}
+ \frac{2}{3} \langle \phi_s^2 \phi_d^2 \rangle 
\left( 4\overline{ |s|^2 |d|^2}
+\overline{s^2 {d^\ast}^2}+\overline{d^2 {s^\ast}^2} \right) , 
\label{eq:a.2}
\end{eqnarray}
where 
\begin{eqnarray}
R_3 &=& \frac{2}{3}\langle \phi_d^4 \rangle |d|^2d -\frac{2}{3}\gamma\xi^2
\langle \phi_d^4 [ 4 \{ (\hat{\bf v}\cdot {\bf q})^2 d \} |d|^2
+d^2 \{ (\hat{\bf v}\cdot {\bf q})^2 d \} ^\ast
-2d|(\hat{\bf v}\cdot {\bf q})d|^2 
+ 3 \{ (\hat{\bf v}\cdot {\bf q})d \} ^2d^\ast ] \rangle, 
\label{eq:a.3} 
\end{eqnarray}
\begin{equation}
R_5 = - \frac{1}{3}\gamma \langle \phi_d^6 \rangle |d|^4 d, 
\label{eq:a.4} 
\end{equation}
\begin{equation}
K = 2 \xi^2 \langle \phi_d^2 (\hat{\bf v}\cdot{\bf q})^2 \rangle
-2 \gamma \xi^4 \langle \phi_d^2 (\hat{\bf v}\cdot{\bf q})^4 \rangle , 
\label{eq:a.5} 
\end{equation}
\begin{equation}
K_{ss} = 2 \xi^2 \langle \phi_s^2 (\hat{\bf v}\cdot{\bf q})^2 \rangle, 
\label{eq:a.6} 
\end{equation}
\begin{equation}
K_{sd} = 2 \xi^2 \langle \phi_s \phi_d (\hat{\bf v}\cdot{\bf q})^2 \rangle. 
\label{eq:a.7}
\end{equation}

From $\partial F / \partial {\bf A}({\bf r})=0$, the expression 
of the current density is reproduced as follows, 
\begin{eqnarray}
{\bf j}({\bf r})
&=&
\xi \nabla \times {\bf h}({\bf r})/(\phi_0/\xi^2)
=
-\frac{\phi_0}{2 \kappa^2 \xi}\biggl\{ \frac{1}{\eta_0^2} \biggl(
 d^\ast \frac{\partial K}{\partial {\bf A}} d 
+s^\ast \frac{\partial K_{ss}}{\partial {\bf A}} s 
+s^\ast \frac{\partial K_{sd}}{\partial {\bf A}} d
+d^\ast \frac{\partial K_{ds}}{\partial {\bf A}} s \biggr)
+ \frac{1}{2 \eta_0^4}
d^\ast \frac{\partial R_3}{\partial {\bf A}} \biggr\}.  
\label{eq:a.8}
\end{eqnarray}
Equation (\ref{eq:a.8}) is equivalent to Eq. (\ref{eq:3.3}), 
if we neglect the term of $d^\ast (\partial R_3 /\partial {\bf A})$  
which is the order $O(|d_0|^4)$. 
From $\overline{d^\ast (\partial F / \partial d^\ast)}=0$ and 
$\overline{s^\ast (\partial F / \partial s^\ast)}=0$, we obtain 
the following relations, 
\begin{eqnarray}
\eta_0^{-2}\overline{|d|^2}
&=&
\eta_0^{-2} \left( 
\overline{d^\ast K d}+\overline{d^\ast K_{sd} s} \right)
+\eta_0^{-4} \left\{ \overline{d^\ast R_3}
+\frac{2}{3}\langle \phi_s^2 \phi_d^2 \rangle 
\left( 2 \overline{|s|^2|d|^2}+ \overline{s^2 {d^\ast}^2} \right) \right\} 
+\eta_0^{-6}\overline{d^\ast R_5} , 
\label{eq:a.9}
\end{eqnarray}
\begin{eqnarray}
-\frac{\overline{|s|^2}}{c_s \gamma \eta_0^2}
&=&
\eta_0^{-2} \left( 
\overline{s^\ast K_{ss} s}+\overline{s^\ast K_{sd} d} \right)
+\eta_0^{-4} \left\{ 
\frac{2}{3}\langle \phi_s^4 \rangle \overline{|s|^4} 
+\frac{2}{3}\langle \phi_s^2 \phi_d^2 \rangle 
\left( 2 \overline{|d|^2 |s|^2} + \overline{d^2 {s^\ast}^2} \right) \right\}. 
\label{eq:a.10}
\end{eqnarray}
By substituting Eqs. (\ref{eq:a.9}) and (\ref{eq:a.10}) into 
Eq. (\ref{eq:a.1}), the free energy is reduced to 
\begin{equation}
\frac{F}{F_0}= 
\kappa^2 \frac{\overline{{\bf H}({\bf r})^2}}{(\phi_0 /\xi^2)^2 } 
-\frac{1}{2} \eta_0^{-4} D_4[d,s] 
-\frac{2}{3} \eta_0^{-6} \overline{d^\ast R_5} . 
\label{eq:a.11}
\end{equation}
By using the magnetic flux density ${\bf B}$ in Eq. (\ref{eq:4.2}), 
\begin{equation}
{\bf H}({\bf r})={\bf H}_0 + {\bf h}({\bf r})
= {\bf B}-\overline{{\bf h}({\bf r})}+{\bf h}({\bf r}). 
\label{eq:a.12}
\end{equation}
Thus, $\overline{{\bf H}({\bf r})^2}$ in the first term of 
Eq. (\ref{eq:a.11}) is written as  
\begin{equation}
\overline{{\bf H}({\bf r})^2}
={\bf B}^2 - {\overline{{\bf h}}}^2 +\overline{{\bf h}^2}. 
\label{eq:a.13}
\end{equation}

We now consider the behavior in the immediate vicinity of $H_{c2}$. 
The quantities $d({\bf r})$,  $s({\bf r})$,  ${\bf A}({\bf r})$ 
and ${\bf H}({\bf r})$ are, respectively,  divided into the values  
at $H_{c2}$ (we denote them as $d_c$, $s_c$, ${\bf A}_c$ and ${\bf H}_{c2}$) 
and the deviations from them, 
\begin{equation}
d=d_c + \delta d, \quad 
s= s_c + \delta s, \quad 
{\bf A}={\bf A}_c + \delta{\bf A}, 
\label{eq:a.14}
\end{equation}
\begin{equation}
{\bf H}({\bf r})={\bf H}_{c2} 
+ \left\{{\bf H}_0 - {\bf H}_{c2} + {\bf h}({\bf r}) \right\}, 
\label{eq:a.15}
\end{equation}
where we have relations, 
\begin{equation}
\nabla\times{\bf A}_c = {\bf H}_{c2}, \quad 
\nabla\times\delta{\bf A}= {\bf H}_0 - {\bf H}_{c2} + {\bf h}({\bf r}). 
\label{eq:a.16}
\end{equation}
By using Eq. (\ref{eq:a.14}), the kernels $K$, $K_{ss}$ and $K_{sd}$ 
are also divided into two parts, for example,  
\begin{equation}
K({\bf A})=K^{(c)}
+ \frac{\partial K^{(c)}}{\partial {\bf A}}\cdot \delta{\bf A}, 
\label{eq:a.17}
\end{equation}
where we write 
\begin{equation}
K^{(c)}= K({\bf A}_c), \quad
\frac{\partial K^{(c)}}{\partial {\bf A}}
= \left. 
\frac{\partial K}{\partial {\bf A}} \right|_{{\bf A}={\bf A}_c} . 
\label{eq:a.18}
\end{equation}
While the deviations $\delta d /d_c$, $\delta s /s_c$ and 
$|\delta{\bf A}|/|{\bf A}_c|$ are in the order of small quantity
$(H_0 - H_{c2})/H_{c2}$, the order parameters $d_c$ and $s_c$ themselves 
are also small quantities in the immediate vicinity of  $H_{c2}$. 
The order parameters $d_c$ and $s_c$, which have been 
already obtained in Sec. \ref{sec:2}, satisfy the linearized GL 
equations, 
\begin{equation}
- d_c + K^{(c)} d_c + K_{sd}^{(c)} s_c =0, 
\label{eq:a.19}
\end{equation}
\begin{equation} 
(c_s \gamma)^{-1} s_c + K_{ss}^{(c)} s_c + K_{sd}^{(c)} d_c =0. 
\label{eq:a.20}
\end{equation}
Equations (\ref{eq:a.19}) and (\ref{eq:a.20}) are equivalent to 
Eqs. (\ref{eq:2.10}) and (\ref{eq:2.11}), 
respectively. 
By substituting Eqs. (\ref{eq:a.14}) and (\ref{eq:a.17}), 
Eq. (\ref{eq:a.8}) is reduced to 
\begin{eqnarray}
{\bf j}({\bf r}) 
&=& 
\xi\nabla \times{\bf h}({\bf r})/(\phi_0/\xi^2)  
=
-\frac{\phi_0}{2 \kappa^2 \eta_0^2 \xi} \biggl(
 d_c^\ast \frac{\partial K^{(c)}}{\partial {\bf A}} d_c
+s_c^\ast \frac{\partial K_{ss}^{(c)}}{\partial {\bf A}} s_c
+d_c^\ast \frac{\partial K_{sd}^{(c)}}{\partial {\bf A}} s_c
+s_c^\ast \frac{\partial K_{sd}^{(c)}}{\partial {\bf A}} d_c \biggr)
\label{eq:a.21}
\end{eqnarray}
within the order $O(|d_0|^2)$. 
In this order, ${\bf j}({\bf r})$ and ${\bf h}({\bf r})$ 
in Eq. (\ref{eq:a.21}) are reduced to 
the ones obtained in Sec. \ref{sec:3} [Eqs. (\ref{eq:3.15}), 
(\ref{eq:3.16}) and (\ref{eq:3.29})].

On the other hand, by substituting Eqs. (\ref{eq:a.14}) 
and (\ref{eq:a.17}) into 
Eqs. (\ref{eq:a.9}) and (\ref{eq:a.10}) and by using 
Eqs. (\ref{eq:a.19}) and (\ref{eq:a.20}), we obtain 
\begin{equation}
\kappa^2 (\phi_0/\xi^2)^{-2} 
\overline{ ( \nabla\times{\bf h}) \cdot \delta{\bf A}} 
= \frac{1}{2}\eta_0^{-4} D_4 ,  
\label{eq:a.22}
\end{equation}
where $\nabla\times{\bf h}$ comes from the relation of 
Eq. (\ref{eq:a.21}). 
From Eqs. (\ref{eq:a.16}) and (\ref{eq:4.2}), there is a relation 
\begin{eqnarray}
\overline{ ( \nabla \times{\bf h}) \cdot \delta{\bf A} }
&=&
\overline{ (  \nabla \times\delta{\bf A} ) \cdot {\bf h}   }
=
\overline{ ({\bf H}_0 - {\bf H}_{c2} + {\bf h})\cdot {\bf h} }
=
({\bf B} - {\bf H}_{c2})\cdot \overline{{\bf h}}
- {\overline{{\bf h}}}^2 + \overline{{\bf h}^2}. 
\label{eq:a.23}
\end{eqnarray}
Thus, by using Eqs. (\ref{eq:a.22}) and (\ref{eq:a.23}), 
we obtain an identity 
\begin{equation}
|d_0|^2 = \frac{\displaystyle \kappa^2 
\frac{({\bf B} - {\bf H}_{c2})\cdot \overline{\bf h} }
{(\phi_0/\xi^2)^2 |d_0|^2} }
{\displaystyle \kappa^2 
\frac{{\overline{\bf h}}^2 - \overline{{\bf h}^2} }
{(\phi_0/\xi^2)^2 |d_0|^4 } 
+ \frac{D_4[d_c,s_c]}{2 \eta_0^4 |d_0|^4} }, 
\label{eq:a.24}
\end{equation}
which determines the amplitude of the order parameter. 
Equation (\ref{eq:a.24}) corresponds to the ``first Abrikosov identity''. 

By substituting Eqs. (\ref{eq:a.13}), (\ref{eq:a.14}) and 
(\ref{eq:a.17}) into Eq. (\ref{eq:a.11}), the free energy 
is written as 
\begin{eqnarray}
\frac{F}{F_0}
&=&
\kappa^2 \frac{{\bf B}^2}{(\phi_0 /\xi^2)2}
- \left\{ \kappa^2 \frac{{\overline{{\bf h}}}^2 -\overline{{\bf h}^2}}
{(\phi_0 /\xi^2)^2 |d_0|^4}
+ \frac{D_4[d_c,s_c]}{2 \eta_0^4 |d_0|^4} \right\} |d_0|^4
+O(|d_0|^6) .
\label{eq:a.25}
\end{eqnarray}
Equation (\ref{eq:4.1}) is obtained by the substitution of 
(\ref{eq:a.24}) into Eq. (\ref{eq:a.25}). 

As for $D_4$ defined in Eq. (\ref{eq:a.2}), the terms except for 
$\overline{d R_3}$ is in the order $O(\gamma^2)$, since $s$ is 
in the order $O(\gamma)$ as is seen from Eq. (\ref{eq:2.26}). 
Thus, $D_4[d_c,s_c]$ of Eq. (\ref{eq:4.4}) is obtained by substituting 
Eqs. (\ref{eq:2.9}) and (\ref{eq:2.25}) into Eqs. (\ref{eq:a.2}) 
and (\ref{eq:a.3}).



\begin{figure}
\caption{
The configuration of the vortex lattice and the coordinate system. 
The vortex centers are shown by solid circle. 
And ${\bf r}_1$ and ${\bf r}_2$ are unit vectors. 
We set the $x$ axis along the direction of ${\bf r}_1$. 
The $a$ axis (or the $b$ axis) forms an angle of $\theta_0$ from the 
$x$ axis (or the $y$ axis). 
}
\label{fig:1}
\end{figure}

\begin{figure}
\caption{
Spatial variation of the induced $s$-wave order parameter for the case 
of a square lattice ($a_y/a_x=0.5$, $\zeta=0.5$). 
(a) Contour plot of the amplitude, $|s({\bf r})/ \gamma s_2^{(1)}|$. 
The region $1.9 a_x \times 1.9 a_x$ is presented. 
The $a$ axis and the $b$ axis are along the horizontal and the vertical 
directions. 
(b) The corresponding position for the vortex of the $s$-wave order 
parameter and its winding number are schematically presented. 
The solid circle shows the vortex center of the dominant 
$d_{x^2-y^2}$-wave order parameter, where the induced $s$-wave 
component has a vortex with winding number $-1$. 
The $s$-wave component has an extra vortex at the point marked x.
The solid line shows the unit cell of the vortex lattice. 
}
\label{fig:2}
\end{figure}

\begin{figure}
\caption{
The same as Fig. 2, but for the case of an oblique lattice 
($a_y/a_x=0.676$, $\zeta=0.5$).
}
\label{fig:3}
\end{figure}

\begin{figure}
\caption{
The equilibrium state of the vortex lattice is schematically presented 
for the case (a). 
Vortices form a shape of an isosceles triangle with $OA=AB$, 
where the base $OA$ is along the $a$ axis.
}
\label{fig:4}
\end{figure}

\begin{figure}
\caption{
Temperature dependence of the unit cell shape of the vortex lattice. 
The ratio $a_y/a_x$ along the curve of $H_{c2}(T)$ 
is presented as a function of $T/T_c$ for 
various mixing of the $s$-wave component, $c_s=-0.2$, $-0.1$, 
0, 0.2, 0.4. 
A thick line shows the pure $d_{x^2-y^2}$-wave case ($c_s=0$). 
The angle $B$ in Fig. 4 gradually varies from $60^\circ$ 
($a_y/a_x=0.866$) to $90^\circ$ ($a_y/a_x=0.5$), 
when temperature is lowered from $T_c$. 
}
\label{fig:5}
\end{figure}

\begin{figure}
\caption{
Spatial variation of the $d_{x^2-y^2}$-wave order parameter. 
Contour plot of the amplitude, $|d({\bf r})/d_0|$, is presented. 
(a) The case of a square lattice ($a_y/a_x=0.5$, $\zeta=0.5$) with 
an equilibrium orientation $\theta_0=0^\circ$ for $\gamma=0.2$. 
(b) The case of a square lattice with an unstable orientation 
$\theta_0=45^\circ$ for $\gamma=0.2$. 
(c) The case of an oblique lattice ($a_y/a_x=0.676$, $\zeta=0.5$) with
an equilibrium orientation $\theta_0=0^\circ$ for $\gamma=0.115$. 
The region $1.9 a_x \times 1.9 a_x$ is presented, where one of the 
vortex centers locates at the center of the figure. 
The $a$ axis and the $b$ axis are along the horizontal and the vertical
directions.
}
\label{fig:6}
\end{figure}

\begin{figure}
\caption{
Spatial variation of the current. 
Contour plot of the amplitude, 
$|{\bf j}({\bf r})|/(\pi \epsilon^{1/2} \xi /2 \kappa^2 \eta^2 )$, 
is presented. 
(a), (b) and (c) correspond to the cases of Fig. 6 (a), (b) and (c), 
respectively. 
}
\label{fig:7}
\end{figure}

\begin{figure}
\caption{
Spatial variation of the induced magnetic field. 
The contour plot of $h_z({\bf r})/|h_z^{(0)}|$ is presented. 
(a), (b) and (c) correspond to the cases of Fig. 6 (a), (b) and (c), 
respectively. 
}
\label{fig:8}
\end{figure}

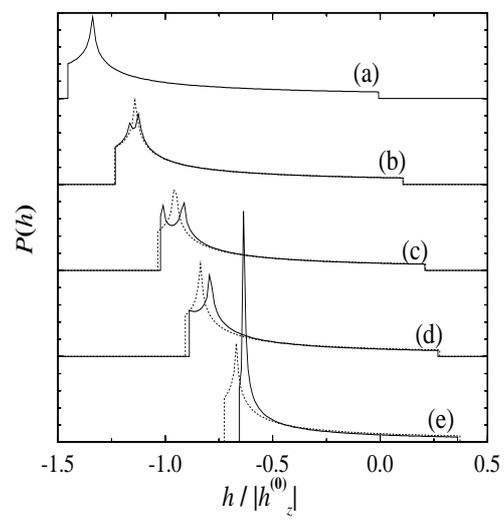
\begin{figure}
\caption{
The magnetic field distribution function $P(h)$ as a function of 
$h/|h_z^{(0)}|$ for various temperatures; 
$\gamma=$0 (a), 0.06 (b), 0.115 (c), 0.15 (d) and 0.2 (e), 
which correspond to $T/T_c=$1, 0.936, 0.881, 0.848 and 0.802, respectively. 
Solid lines are for the case of a $d_{x^2-y^2}$-wave superconductor, 
where (a) is the triangular lattice, (b) and (c) the oblique lattice, 
(d) and (e) the square lattice. 
There, parameters are set as $\zeta=0.5$, $\theta_0=$0 and 
$a_y/a_x=$0.866 (a), 0.799 (b), 0.676 (c), 0.5 (d), 0.5 (e). 
Dotted lines are for the case of an isotropic $s$-wave superconductor, 
where (a)-(e) are the triangular lattice ($\zeta=0.5$, $a_y/a_x=0.866$). 
For clarity, the curves have been sifted along the vertical axis. 
}
\label{fig:9}
\end{figure}

\end{document}